\documentclass[sigconf,nonacm,screen]{acmart}

\setcopyright{none}    
\renewcommand\footnotetextcopyrightpermission[1]{} 

\usepackage{hyperref}
\usepackage{url}
\usepackage{graphicx}
\usepackage{booktabs}
\usepackage{algorithm}
\usepackage{algorithmic}
\usepackage{bbding}
\usepackage{url}
\usepackage{graphicx}
\usepackage{subfig}
\usepackage{wrapfig}
\usepackage{listings}

\usepackage{multirow}
\usepackage{multicol}
\usepackage{makecell}
\usepackage{enumitem}

\hypersetup{colorlinks=true,
linkcolor=cyan,citecolor=cyan,urlcolor=cyan
}

\newcommand{\latinabbrev}[1]{\textit{#1}}
\newcommand{\eg}{\latinabbrev{e.g.}}

\newcommand{\etc}{\latinabbrev{etc}}
\newcommand{\ie}{\latinabbrev{i.e.}}
\newcommand{\wrt}{\latinabbrev{w.r.t.}}

\newcommand{\fname}{\texttt{MetamatBench}}
\newcommand{\interfalceurl}{\url{http://zhoulab-1.cs.vt.edu:5550}}
\newcommand{\codebaseurl}{\url{https://github.com/cjpcool/Metamaterial-Benchmark}}

\theoremstyle{plain}
\newtheorem{definition}{Definition}

\renewcommand{\paragraph}[1]{%
 \par\smallskip
\noindent\textbf{\textit{#1}}.\quad%
}

\begin{document}

\title{\fname: Integrating Heterogeneous Data, Computational Tools, and Visual Interface for Metamaterial Discovery}


\author{%
  Jianpeng Chen\textsuperscript{1}, Wangzhi Zhan\textsuperscript{1}, Haohui Wang\textsuperscript{1}, Zian Jia\textsuperscript{2,5}, Jingru Gan\textsuperscript{3}, Junkai Zhang\textsuperscript{3}, Jingyuan~Qi\textsuperscript{1}, Tingwei Chen\textsuperscript{1}, Lifu Huang\textsuperscript{4}, Muhao Chen\textsuperscript{4}, Ling Li\textsuperscript{5}, Wei Wang\textsuperscript{3}, Dawei Zhou\textsuperscript{1}
}
\affiliation{%
\institution{
\textsuperscript{1}Virginia Tech;
\textsuperscript{2}Princeton University;
\textsuperscript{3}University of California, Los Angeles;
\textsuperscript{4}University of California, Davis;
\textsuperscript{5}University of Pennsylvania
}
\country{}
}

\renewcommand{\shortauthors}{J. Chen, W. Zhan, et al.}

\begin{abstract}
Metamaterials, engineered materials with architected structures across multiple length scales, offer unprecedented and tunable mechanical properties that surpass those of conventional materials. However, leveraging advanced machine learning (ML) for metamaterial discovery is hindered by three fundamental challenges: {(C1) Data Heterogeneity Challenge} arises from heterogeneous data sources, heterogeneous composition scales, and heterogeneous structure categories; {(C2) Model Complexity Challenge} stems from the intricate geometric constraints of ML models, which complicate their adaptation to metamaterial structures; and {(C3) Human-AI Collaboration Challenge} comes from the ``dual black-box'' nature of sophisticated ML models and the need for intuitive user interfaces. 
To tackle these challenges, we introduce a unified framework, named {\fname}, that operates on three levels. 
(1) At the \emph{data level}, we integrate and standardize 5 heterogeneous, multi-modal metamaterial datasets.
(2) The \emph{ML level} provides a comprehensive toolkit that adapts 17 state-of-the-art ML methods for metamaterial discovery. It also includes a comprehensive evaluation suite with 12 novel performance metrics with finite element-based assessments to ensure accurate and reliable model validation.
(3) The \emph{user level} features a visual-interactive interface that bridges the gap between complex ML techniques and non-ML researchers, advancing property prediction and inverse design of metamaterials for research and applications.
\fname\ offers a unified platform deployed at \interfalceurl\ that enables machine learning researchers and practitioners to develop and evaluate new methodologies in metamaterial discovery. 
For accessibility and reproducibility, we open-source our benchmark and the codebase at \codebaseurl.
\end{abstract}

\begin{teaserfigure}
\centering
\vspace{-1em}
  \includegraphics[width=\textwidth]{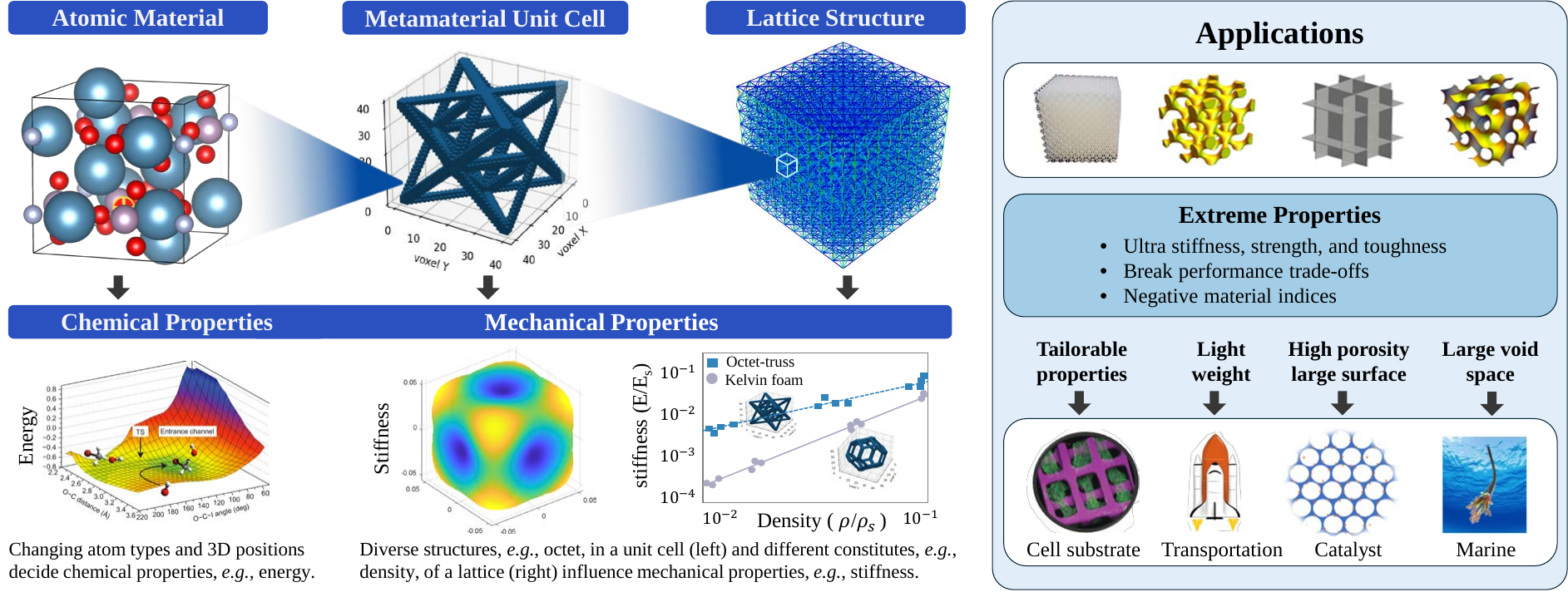}
    \put(-424,22){\tiny{\cite{otto2012single}}}
    \put(-64,179){\tiny{\cite{jia2020engineering}}}
    \vspace{-1em}
  \caption{
An overview of metamaterials. Metamaterials are microstructured materials with effective material properties beyond their compositions. The multiscale architecture of metamaterial enables high mechanical efficiency and unusual properties,  showing great potential in applications like biomedical devices, transportation systems, robotics, \etc.  
  }
  \label{fig:metamat}
\end{teaserfigure}

\maketitle

\section{Introduction}

Metamaterials are an emerging class of materials that achieve unusual material properties through designed architecture at multiple length scales. They have been extensively studied in the past decades for their superior, tunable, and programmable material properties, demonstrating huge potential in diverse applications~\cite{paul2010optical,engheta2006metamaterials}.
As illustrated in Figure~\ref{fig:metamat} (left), while conventional material properties are dominated by their atomic structure~\cite{yao2018engineering,otto2012single}, architected metamaterials emphasize the structure of a material, with scales ranging from the nano- and micro-scales up to macro-scale. Metamaterials, \eg, truss-based metamaterials, studies often focus on designing the geometric structure of a unit cell. 
Micro-lattice materials have shown high stiffness, damage tolerance, reconfigurable and programmable properties, and even negative material indices (such as negative Poisson's ratio and negative thermal conductivity)~\cite{jia2020engineering}. Such high performance with unusual properties drives the wide application of metamaterials in various engineering fields (lightweight metamaterial empowers space transport systems, metamaterials with large void space enhance marine application, low thermal conductivity metamaterials are applied to thermal protection systems, \etc.), as shown in Figure~\ref{fig:metamat} (right).

Because of their truss-based geometric characteristics, metamaterials are typically modeled as 3D graphs composed of nodes and edges to study how their 3D structures influence mechanical properties. 
Existing works~\cite{mace_ve,mCGCNN,uniftruss} generally employ graph neural networks (GNNs)~\cite{gcn,vgae,VGMGC} to capture the structural information for mechanical property prediction~\cite{indurkar2022predicting,mCGCNN} or metamaterial inverse design~\cite{uniftruss,Modulus}.
In parallel, advanced studies in geometric machine learning (ML) have extensively explored techniques to integrate rich 3D structural information in molecules~\cite{PotentialNet,schnet,spherenet,gasteiger2021gemnet,EquiformerV2,satorras2021n,atz2021geometric,MACE,visnet,chen20243d} and crystals researches ~\cite{CDVAE,SyMat,diffcsp,CGCNN,zeni2025generative}. Although many studies have benchmarked these methods on atomic crystal and molecular scales~\cite{MatBench,M2Hub,Gom3D,omat24}, it remains unclear how advanced geometric ML approaches perform in the multiscale metamaterial domain. Consequently, there is a critical need to establish a standardized benchmark and evaluation platform that bridges the gap between current metamaterial modeling techniques and advanced geometric ML approaches, ensuring comprehensive integration and assessment of 3D structural information in metamaterial discovery.
 

In this paper, we identify three key challenges in constructing a metamaterial benchmark.
\textbf{C1:~Data heterogeneity.} This challenge arises from (1) diverse data sources, \eg, structural geometry, mechanical properties, and experimental measurements, (2) complex structure categories, \eg, trusses, shells, foams, \etc., and (3) rich properties \eg, stiffness and modulus. 
\textbf{C2:~Model complexity.} The second challenge stems from the complexity of ML models and their potential incompatibility with multiscale metamaterials. Specifically, these ML models typically have a complex taxonomy with various backbones and geometric constraints. Additionally, they generally target atomic graphs and chemical properties, which are typically incompatible with metamaterials. These complexities pose challenges to evaluating and benchmarking advanced ML models on metamaterial applications. How to integrate and evaluate these complex ML models is a problem of pressing needs.
\textbf{C3:~Human-AI collaboration.} 
A goal of this work is to empower metamaterial researchers to easily leverage advanced ML models to accelerate progress in related fields. Achieving this requires fostering effective human-AI collaboration across diverse research domains.
On one hand, sophisticated ML models often function as black box for researchers who may lack expert knowledge about advanced ML models. Instead, a visual-interactive interface may help researchers interact with the AI system.
On the other hand, human users are also black box for the AI system. It is hard for the AI system to anticipate how researchers use it.
To address this ``dual black-box'' challenge, a \textit{human-AI collaborative} interface is essential to facilitate metamaterial research.

In this paper, we propose \fname\ as shown in Figure~\ref{fig:overview}, which contributes to the data level, ML level, and user level, establishing a robust, comprehensive, and user-friendly benchmark system.
Within \fname, the data level combines heterogeneous and multi-modal data sources into a unified framework to tackle the first challenge.
The intermediate ML level consists of a model toolbox and an evaluation toolbox. The model toolbox focuses on two fundamental tasks, and assembles a wide range of ML models with various geometric characteristics for a comprehensive comparison. The evaluation toolbox employs a multi-perspective evaluation framework with several novel metrics to ensure robust evaluations.
The topmost user level emphasizes human-AI collaboration, mitigating the dual black-box issue by providing a visual-interactive interface.
This three-level system is integrated to advance the exploration and research of metamaterials. To the best of our knowledge, \fname\ is the first benchmark for metamaterial that integrates heterogeneous data, ML models, novel metrics, and visual-interactive interface.
The overall contributions can be summarized as follows.
\begin{itemize}[leftmargin=*, topsep=-1mm]
    \item \textbf{Database Development:} We collect and process five metamaterial datasets covering multi-modal lattice structures, and unify the representation of three 3D graph metamaterial datasets. 
    \item \textbf{ML Toolbox Development:} We introduce a \textit{model toolbox} that integrates 17 models designed for 3D crystal materials and molecules, and adapts them to metamaterial learning to assess their effectiveness in metamaterial tasks.
    Moreover, we develop a \textit{evaluation toolbox} that includes an evaluation framework with novel metrics for robust metamaterial assessment and finite element (FE) computation-based metrics for physics-aware evaluation.
    \item \textbf{Visual-Interactive Interface Development:} We design a visual-interactive interface to facilitate human-AI collaboration and data visualization. This interface helps metamaterial researchers explore advanced methods and choose appropriate ML models, thereby bridging the gap between metamaterial research and machine learning.
    We deploy the interface at \interfalceurl.
    \item \textbf{Open-Sourced Codebase:} For accessibility and reproducibility, we have open-sourced our benchmark and the codebase at \codebaseurl.
\end{itemize}
\vspace{-1em}
\section{Preliminary}
\subsection{Previous Benchmarks}
\label{sec:pre_benchamrks}
In recent years, many benchmarks of ML for scientific discovery have emerged with the breakthroughs made in AI for science~\cite{jumper2021highly,zeni2025generative,wang2024ab}. These benchmarks have explored the application of complex ML to 3D atomic-scale scientific discovery. However, they generally focus on crystal or molecular materials with chemical properties, as demonstrated in the top part of Table~\ref{tab:compData}. For instance, \cite{pickard2020airss,castelli2012new,castelli2012computational,MP,MatBench,OC20,rosen2022high} focus on benchmarking crystal materials where atom type and chemical properties dominate the learning space, and  \cite{QM9,OMDB} benchmark ML models on molecular space that is also based on atom type and chemical properties. The metamaterials that specifically focus on multiscale architectures and mechanical properties still lack exploration. Therefore, to fill the gap, a comprehensive metamaterial benchmark with a unified representation is needed.

\vspace{-0.5em}
\subsection{Unified Metamaterial Representation}\label{sec:pre_metarep}
\begin{figure}[tb]
    \centering
    \includegraphics[width=\linewidth,height=16.5em]{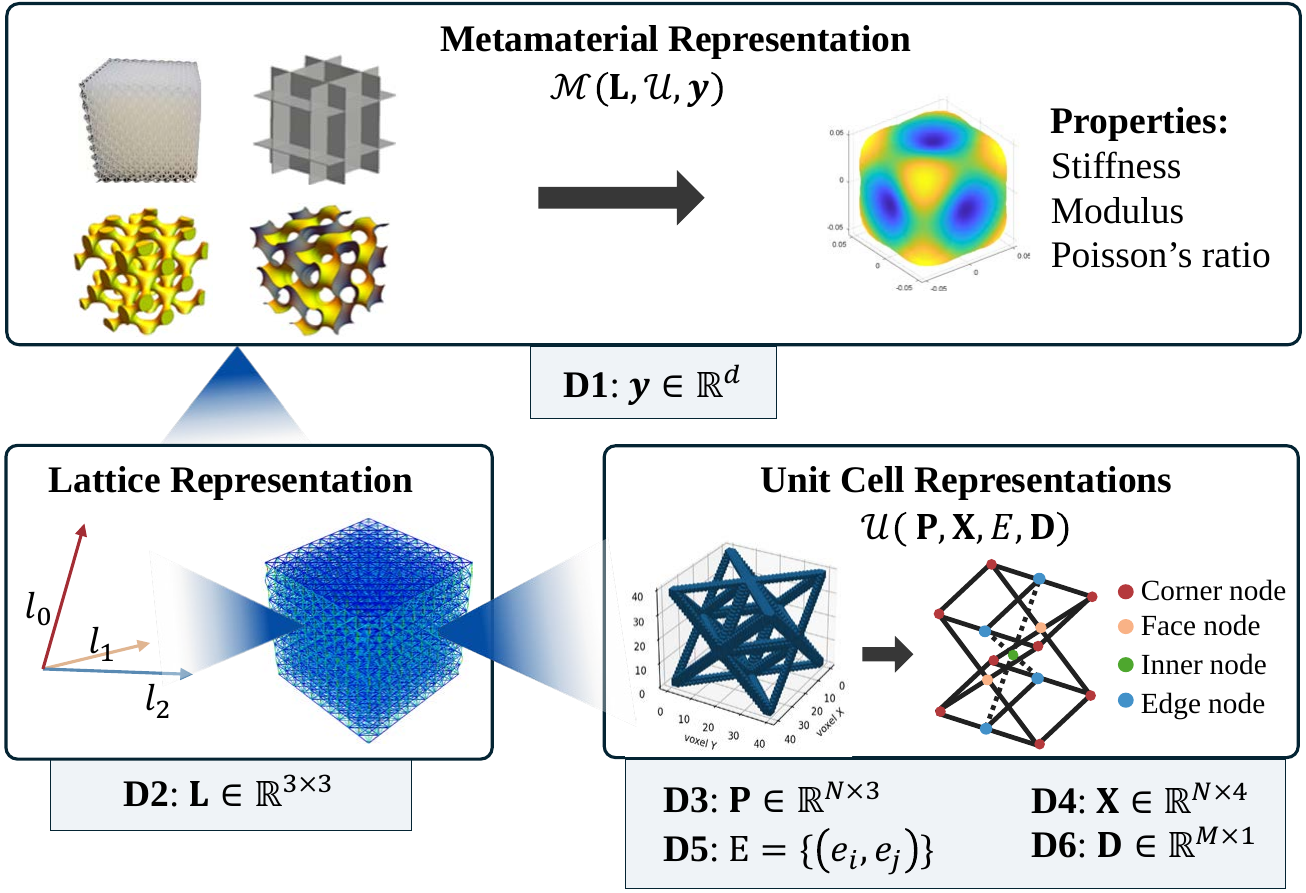}
    \vspace{-2em}
    \caption{Unified metamaterial representation.}
    \label{fig:uniRep}
\end{figure}
To address data heterogeneity (C1), we introduce a metamaterial representation to unify 3D graph datasets and benchmark them with advanced geometric graph learning methods. This unified metamaterial representation considers a six-dimensional learning space, de-emphasizing atomic element information while highlighting structural lattice characteristics. 
In general, we denote $\mathcal{M}(\mathbf{L}, \mathcal{U}, \mathbf{y})$ as a metamaterial. Figure~\ref{fig:uniRep} illustrates the hierarchical six-dimensional learning space of a metamaterial considering the necessary details of \textit{unit cell} and \textit{lattice} for \textit{metamaterial} applications. The six dimensions of the learning space are labeled D1 through D6 for reference:
\vspace{-0.5em}
\begin{definition}[Metamaterial Property]
The mechanical properties of a metamaterial (D1) are represented as $\mathbf{y} \in \mathbb{R}^{d}$, where $d$ is the property dimension.
\end{definition}
\vspace{-0.5em}
\begin{definition}[Lattice Representation]
The metamaterial lattice structure (D2) is denoted by $\mathbf{L} = [\mathbf{l}_0, \mathbf{l}_1, \mathbf{l}_2]^\mathrm{T} \in \mathbb{R}^{3 \times 3}$, where $\mathbf{l}_d \in \mathbb{R}^3$, capturing the periodic angles and lattice lengths in 3D space.
\end{definition}
\vspace{-0.5em}
\begin{definition}[Unit Cell Representation]
The metamaterial unit cell is represented by $\mathcal{U}(\mathbf{P}, \mathbf{X}, E, \mathbf{D})$, composed of four components: node coordinates (D3), node attributes (D4), edge connections (D5), and Edge attributes (D6).
\end{definition}
\noindent To be specific, \textbf{node coordinates (D3)} denote the $N$ node positions in 3D Cartesian system: $\mathbf{P} = [\mathbf{p}_0, \mathbf{p}_1, \ldots, \mathbf{p}_{N-1}]^\mathrm{T} \in \mathbb{R}^{N \times 3}$, where $\mathbf{p}_i \in \mathbb{R}^3$. In addition, we provide the transformed fractional coordinates in the unified representation for convenient computation.
\textbf{Node attributes (D4)} $\mathbf{X} \in \mathbb{R}^{N \times 4}$ denotes the specifically designed one hot encoding of four types of $N$ nodes, \ie, face node, corner node, edge node, and inner node, as depicted in Figure~\ref{fig:uniRep}. 
Unlike atomic graphs where node attributes naturally depict the element type, the designed representation of node attributes emphasizes structural information. Specifically, similar to \cite{mace_ve}, we classify all nodes into outer nodes and inner nodes, where inner nodes are the nodes inside the unit cell while outer nodes are shared by multiple unit cells in a lattice since the periodical repetition, for example, face nodes are shared by two unit cells, edge nodes are shared by four unit cells, and corner nodes are shared by eight unit cells.
\textbf{Edge connections (D5)} $E = \{(e_i, e_j)\}$ describe the edges between $i$-th and $j$-th node.
\textbf{Edge attributes (D6)} $\mathbf{D}\in\mathbb{R}^{M\times1}$ record the auxiliaries for $M$ edges, \eg, edge diameter which determines the density of a unit cell.
\vspace{-0.5em}
\section{\fname~Development}
\subsection{\fname: Overview}
\begin{figure*}[tb]
    \centering
        \includegraphics[width=0.91\linewidth]{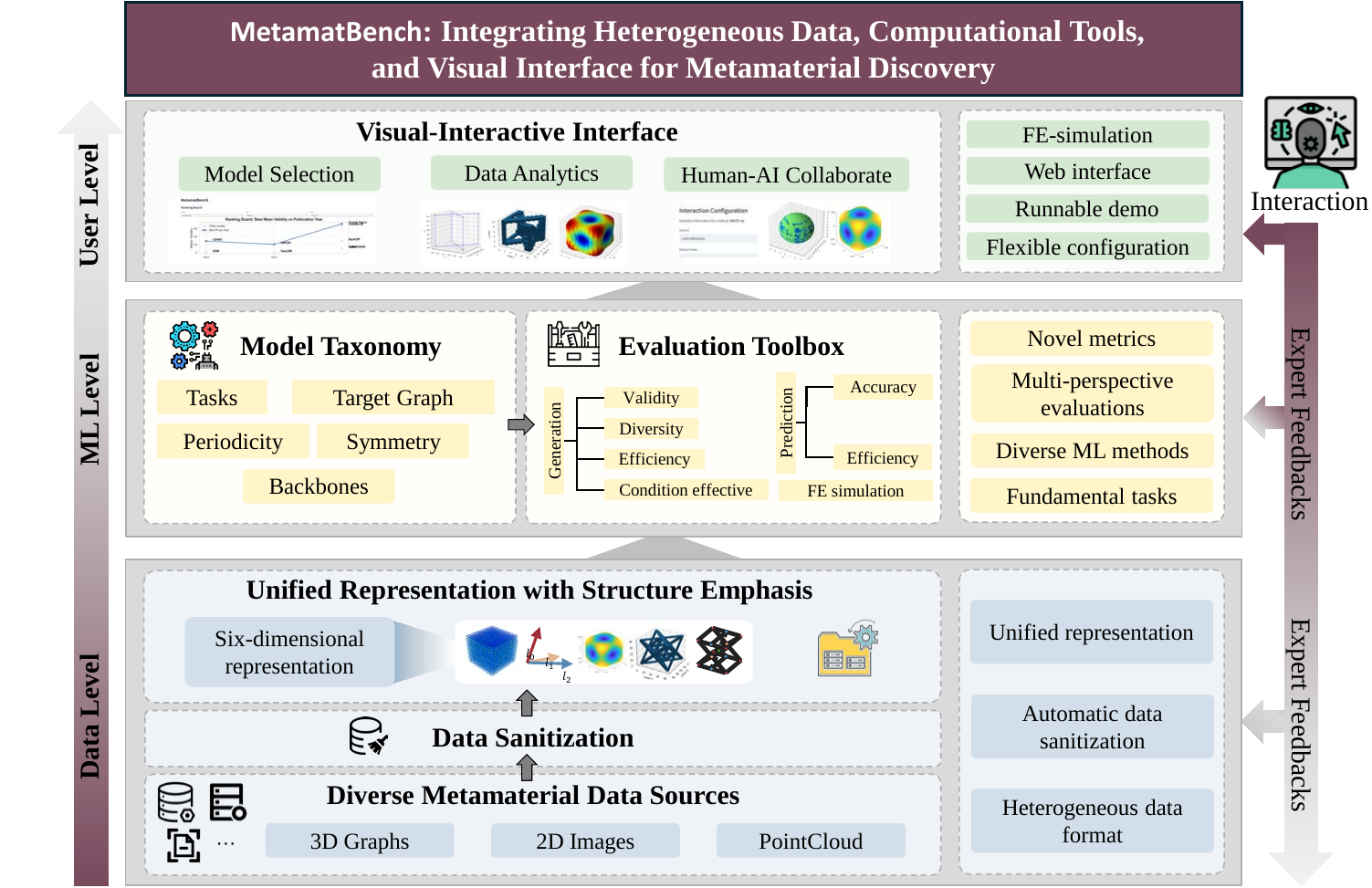}
    \caption{An overview of \fname: bottom dataset level provides a base for various metamaterial applications; middle ML level includes ML models and evaluation toolbox benefits researchers in finding the best suitable models; top user level visual-interactive interface enables human-AI dual black-box collaborations.}
    \vspace{-1em}
    \label{fig:overview}
\end{figure*}

\fname~is a multi-level system containing data level, ML level, and user level, as illustrated in Figure~\ref{fig:overview}.
In this section, we introduce the development of \fname~from bottom to top, \ie, from database development (Section~\ref{sec:database_dev}), ML model toolbox and evaluation toolbox development (Section~\ref{sec:ML_dev}), to visual-interactive interface development (Section~\ref{sec:interface_dev}).
\vspace{-1em}
\subsection{\fname: Database Development}\label{sec:database_dev}
The data level aims to mitigate data heterogeneity challenges (C1) by collecting and preprocessing various data sources to a unified data representation.
Overall, Table~\ref{tab:compData} compares the datasets included in \fname\ (bottom portion of the table) to those commonly used in prior work (top portion). It shows that \fname\ includes a large number of datasets with a specific focus on metamaterial and mechanical properties. \fname\ collects heterogeneous metamaterial datasets across multiple modalities. 
We then apply a unified data sanitization process and use the unified metamaterial representation (defined in Section~\ref{sec:pre_metarep}) for the integrated 3D graph datasets. This unified process enables fair evaluation of different 3D graph models, and it also paves the way for future exploration with multiple data modalities for metamaterial design.
\paragraph{Heterogeneous Data Sources}
Metamaterials often require complicated structures to achieve desired properties, which can be difficult to design and predict. Therefore, we anticipate that the integration of multi-modal data will alleviate this challenge and further advance the field, attracting more researchers to explore this topic. We collect three modalities of datasets in this benchmark, including three 3D graph-based datasets (MetaStiffness~\cite{Stiffness}, MetaModulus~\cite{Modulus}, and MetaTruss~\cite{uniftruss}), one 2D image dataset (LagrangianFrame~\cite{bastek2023inverse}), and one point cloud dataset (PointCloud~\cite{chan2021metaset}). To be specific,
\emph{MetaModulus~\cite{Modulus}} is a metamaterial dataset constructed from two publicly accessible crystal material databases, RCSR~\citep{RCSR} and EPICNET~\citep{EPINET}. This dataset provides three properties, \ie, Young's modulus, Shear modulus, and Poisson's ratio.
\emph{MetaStiffness~\cite{Stiffness}} is a dataset that employs seven fundamental lattices to construct metamaterials. By combining three lattice types, 262 unique topologies are generated, with additional variants produced through rotation and scaling transformations. Each structure's stiffness tensor is characterized by 21 independent elastic constants.
\emph{MetaTruss~\cite{uniftruss}} is generated from five elementary truss lattices, randomly adding or deleting edges and adjusting the node positions. It filters out the physically invalid lattices and randomly selects lattices to construct the final dataset.
\emph{LagrangianFrame~\cite{bastek2023inverse}} proposes representing metamaterials in a Lagrangian frame (instead of the traditional Eulerian frame) and provides a 2D dataset of metamaterial structures generated in this Lagrangian coordinate system.
\emph{PointCloud~\cite{chan2021metaset}} produces 29,400 3D point-cloud data constructed by sampling existing datasets to avoid inherent data biases.
\paragraph{Data Sanitization}
We observe several issues (described below) on these datasets, leading to the collected datasets not being directly applicable for benchmarking. Therefore, we design a multi-perspective prototype following previous works~\cite{uniftruss,Stiffness} to filter out invalid structures for these datasets. This prototype includes following hard constraints to automatically filter out invalid structures, and it is also considered in our proposed evaluation toolbox for validity evaluation.
\begin{itemize}[leftmargin=*, topsep=-0.4mm]
    \item \textbf{Metamaterial-Oriented Sanitization}: At the metamaterial level, we observe that some property values are missed in the datasets. For example, many lattices in MetaModulus dataset miss the property of Poisson's ratio. Therefore, these samples with missing values are filtered out to ensure data completeness.
    \item \textbf{Lattice-Oriented Sanitization}: At the lattice level, we discover that the node coordinates might exceed the lattice range, \eg, some nodes in MetaModulus. In addition, we find that some nodes in MetaModulus dataset are extremely close. Therefore, we propose \textit{distance restriction} to filter out these lattices with dispersed or clustered nodes. The lattices containing node distances larger than lattice lengths or smaller than a specified threshold are identified and subsequently removed from the dataset.
    \item \textbf{Unit-Cell-Oriented Sanitization}: At the unit cell level, we find that some structures are invalid. For instance, many unit cells in MetaModulus are physically invalid. We propose that the unit cell (node connection patterns) of metamaterials should satisfy: (1) \textit{Connection graph}: all structures should be connected graphs. (2) \textit{Dangling restriction}: there is no dangling node in a structure, \ie, all nodes have at least two edges connecting to other nodes.
\end{itemize}
After conducting these hard constraints for sanitization, the heterogeneity of three collected datasets is reduced (as analyzed in Section~\ref{sec:analy_data}). The final data statistics are summarized in the bottom rows of Table~\ref{tab:compData}. More statistics can be found in Appendix~\ref{app:dataset}.
\begin{table*}[tb]
\small
\centering
\vspace{-1em}
\caption{Comparison of material datasets. The upper group includes conventional atomic-scale materials commonly used in ML. The lower five datasets, collected by \fname, focus on metamaterials with specialized mechanical properties. }
\vspace{-1em}
\begin{tabular}{l|c|c|c|c}
\Xhline{2\arrayrulewidth}
Dataset  & Design Target & Periodic & Property &  \# Samples \\
\hline
Carbon24~\cite{pickard2020airss} & Atomic-scale crystalline materials  &  \Checkmark & Energy  & 10,153  \\
Perov5~\cite{castelli2012new,castelli2012computational} & Perovskite-type crystalline materials & \Checkmark & Energy  & 18,928   \\ 
MP20~\cite{MP} & Crystalline atomic materials & \Checkmark & Chemical properties  & 45,231 \\
MatBench~\cite{MatBench} & Inorganic bulk materials & \Checkmark  & Chemical properties   & 312–132,752 \\
OC20~\cite{OC20} & Bulk-adsorbate interface materials  & \XSolidBrush  & Energetic &  640,081 \\
QMOF~\cite{rosen2022high} & Metal–organic framework (MOF) materials  & \Checkmark  & Chemical properties & $>$20,000  \\
QM9~\cite{QM9} & Molecular compounds  & \XSolidBrush  & Chemical properties  & $\sim$134,000 \\
OMDB~\cite{OMDB} & Organic crystalline materials  & \Checkmark  & Electronics  & 12,500  \\
\hline
{MetaModulus}~\cite{Modulus} & Architected truss metamaterials for modulus design  & \Checkmark & Three mechanical properties  & 16,707 \\
MetaStiffness~\cite{Stiffness} & Architected truss metamaterials for stiffness optimization  & \Checkmark  & Elastic constants   & 1,048,575 \\
MetaTruss~\cite{uniftruss} & Architected truss metamaterials for homogeneous stiffness  & \Checkmark & Homogeneous stiffness    & 965,736 \\
PointCloud~\cite{chan2021metaset}  & Truss metamaterials (3D point-cloud representation)  & \Checkmark  & Mechanical properties  &  29,400 \\
LagrangianFrame~\cite{bastek2023inverse} & Shell and truss metamaterials (2D Lagrangian frame)   & \Checkmark  & Stress and strain  &  53,007 \\
\Xhline{2\arrayrulewidth}
\end{tabular}
\label{tab:compData}
\end{table*}
\begin{table*}[t]
\small
\vspace{-1em}
    \caption{The statistics of comparison methods. * indicates conditional generation support. Abbreviations: Trans Inv. (Translation Invariance), Glob (Global), Equiv. (Equivariant), Rot. (Rotation), VAE (Variational Auto Encoder), Diff (Diffusion), MPNN (Message Passing Neural Networks), GCN (Graph Convolutional Networks), LatDiff (Latent Diffusion), Perm. (permutation).}
    \vspace{-1em}
    \label{tab:comparisonModel}
    \begin{tabular}{l|ccccc}
    \Xhline{2\arrayrulewidth}
    Methods   & Task  & Design Target & Periodicity  & Symmetry            & Backbone\\
    \hline
    EDM*~\cite{EDM}             & Generation  & Molecule & N/A                       & Equiv.               & Diff\\
    GeoLDM*~\cite{GEOLDM}       & Generation  & Molecule & N/A                       & Equiv.               & LatDiff\\
    DiffCSP~\cite{diffcsp}       & Generation  & Crystal  & Trans Inv.                & Equiv. (lacks lattice Perm. Equiv.)   & Diff\\
    CDVAE~\cite{CDVAE}         & Generation  & Crystal  & Trans Inv.                & Inv Enc + Equiv Dec   & VAE+Diff\\
    EquiCSP~\cite{EquiCSP}       & Generation  & Crystal  & Trans Inv. + Perm Eq.      & Equiv.               & Diff\\
    Cond-CDVAE*~\cite{condCDVAE} & Generation  & Crystal  & Trans Inv.                & Inv Enc + Equiv Dec   & VAE+Diff\\
    SyMat~\cite{SyMat}         & Generation  & Crystal  & Trans Inv.                & Inv Enc + Inv Dec     & VAE+Diff\\
    Crystal-Text-LLM*~\cite{gruver2024finetuned} & Generation  & Crystal  & N/A                       & N/A                  & GPT-2\\
    CrystaLLM*~\cite{CrystaLLM}   & Generation  & Crystal  & N/A                       & N/A                  & LLaMA-2\\
    \hline
    SchNet~\cite{schnet}        & Prediction & Molecule & N/A                       & Glob Inv             & MPNN\\
    SphereNet~\cite{spherenet}   & Prediction & Molecule & N/A                       & Glob Inv             & MPNN\\
    Equiformer~\cite{equiformer}  & Prediction & Molecule & N/A                       & Equiv.               & MPNN\\
    ViSNet~\cite{visnet}         & Prediction & Molecule & Trans. + Rot. Inv             & Equiv.               & MPNN\\
    CGCNN~\cite{CGCNN}           & Prediction & Crystal  & Trans Inv.                & Glob. Inv.             & GCN\\
    ALIGNN~\cite{ALIGNN}         & Prediction & Crystal  & Trans Inv.                & Glob Inv             & GCN\\
    UniTruss~\cite{uniTruss}     & Prediction & Metamaterial & N/A                    & N/A                  & VAE\\
    MACE+ve~\cite{mace_ve}        & Prediction & Metamaterial & N/A                    & Equiv.               & MPNN\\
    \Xhline{2\arrayrulewidth}
    \end{tabular}
    \vspace{-1.2em}
\end{table*}
\subsection{\fname: ML Toolbox Development}\label{sec:ML_dev}
The ML level development aims to integrate and evaluate advanced ML levels for metamaterials applications by addressing the model complexity challenge (C2). In the ML toolbox, we assemble 17 state-of-the-art ML models (Table~\ref{tab:comparisonModel}) into a model toolbox and propose a comprehensive evaluation toolbox with 12 novel metrics (Table ~\ref{tab:evaluation_toolbox}) to evaluate ML model's effectiveness on metamaterial applications.
\paragraph{Model Toolbox} Our model toolbox focuses on two fundamental tasks (\ie, metamaterial generation and property prediction) and assembles a wide range of ML models with various geometric characteristics for a comprehensive comparison. Specifically, to compare 3D graph models for metamaterials, we consider both emerging 3D crystal graph models and the well-developed 3D molecule graph models. These two research areas focus on different aspects of 3D graphs while still sharing similarities for benchmarking metamaterial-based tasks. For example, crystal-targeted models should preserve periodic symmetries due to the periodic nature of the material~\cite{luo2024towards}, which is not necessary for molecules. Instead, the latter requires completeness for distinguishing molecule chirality~\cite{keriven2019universal,SatorrasHW21}. In addition, we specifically compare the performance of two mainstream types of 3D graph models, \ie, equivariant and invariant models~\cite{batzner20223}, on metamaterial datasets.

Overall, to benchmark 3D graph models on metamaterial, the proposed taxonomy in Figure~\ref{fig:overview} covers: (1) generative models and predictive models, (2) crystal material, molecular, and metamaterial graph models, (3) models with different periodicity constraints, (4) models with symmetry constraints, such as equivariant model and invariant model, and (5) models with various backbones. The detailed model taxonomy is summarized in Table~\ref{tab:comparisonModel}.

Based on this taxonomy, we benchmark the two fundamental tasks, \ie, prediction task and generation task. Both tasks are crucial in measuring the effectiveness of ML models for metamaterial learning in various application scenarios.


\vspace{-0.5em}
\paragraph{Evaluation Toolbox}\label{sec:eval_dev}
\begin{table*}[ht]
\small
\centering
\vspace{-1em}
\caption{Overall framework of the proposed evaluation toolbox. $N_L$ is the number of generated structures.}
\vspace{-1em}
\label{tab:evaluation_toolbox}
\begin{tabular}{p{0.09\linewidth}| p{0.14\linewidth}| p{0.7\linewidth}}
\Xhline{2\arrayrulewidth}
Task & Perspective & Metric\\
\hline
\multirow[c]{22}{*}{\rotatebox{0}{\parbox{1.5cm}{Generation}}}
 &  \multirow[c]{9}{*}{Validity}  
 &  $\bullet$ \textbf{Dangling Restriction (Node Level)}:
    $ \mathcal{V}_{DR} = 1 - \frac{N_{D}}{N_L}$, where $N_{D}$ is the number of structures that contain dangling node.\\
 &  &  $\bullet$ \textbf{Connectivity (Edge Level)}:
    $ \mathcal{V}_{C} = \frac{N_{C}}{N_L}$, where $N_{C}$ is the number of structures that are connected graph.\\
 &  &  $\bullet$ \textbf{Symmetry (Unit Cell Level)}:
    $ \mathcal{V}_{S} = \frac{1}{N_L} \sum_{k=1}^{N_L} \frac{N_{S_k} \cdot \sum_{i=1}^{N_k} s_{\mathrm{degree}_i}}{(N_k)^2}$, where $N_k$ is the node number of $k$-th structure, and $N_{S_k}$ is the number of Symmetrical Node that is defined in Definition.~\ref{def:symmetric_node} in $k$-th structure, and $s_{degree_i}$ denotes Symmetry Degree that is defined in Definition~\ref{def:symm_degree}. \\
 &  &  $\bullet$ \textbf{Periodicity (Lattice Level)}:
    $ \mathcal{V}_{P} = \frac{N_P}{N_L}$, where $N_P$ denotes the number of generated structures that satisfy Definition~\ref{def:period}.\\
\cline{2-3}
 & \multirow[c]{4}{*}{Diversity} 
 &  $\bullet$ \textbf{Coverage Recall}:
    $ \text{COV}_R=\frac{1}{N_t}\vert \{i \in [1,\ldots,N_t]: \exists k \in [1,\ldots,N_L],
    D(\mathbf{P}^*_i, \mathbf{P}_k) < \epsilon_{cov} \} \vert$, where $D(\cdot)$ denotes structural distance. Given an error bar $\epsilon_{cov}$, $N_t$ test structures, and node positions $\mathbf{P}_i$ and $\mathbf{P}^*_j$ of $i$-th and $j$-th structures.\\
 &  &  $\bullet$ \textbf{Coverage Precision}:
    $\text{COV}_P=\frac{1}{N_L}\vert \{i \in [1,\ldots,N_L]: \exists k \in [1,\ldots,N_t],
    D(\mathbf{P}_i, \mathbf{P}^*_k) < \epsilon_{cov} \} \vert. $\\
\cline{2-3}
 & \multirow[c]{6}{*}{\parbox{2cm}{Conditional\\Effectiveness\\(Figure~\ref{fig:cond_eff})}}
 & Conditional effectiveness is the mean of all $N_L$ distances: $\frac{1}{N_L}\sum_i^{N_L} \min_{j \in {1,\ldots, K}}\text{Dist}(\mathbf{y}_i, \mathbf{y}_{i,j})$, where Dist is euclidean distance. The $\{\mathbf{y}_i\}_{i=1}^{N_L}$ and $\{\mathbf{y}_i,j\}_{j=k}^{K}$ are obtained by four steps.\\
 &  &  \textbf{Step 1}: Generate $N_L$ lattices conditioned on $N_L$ properties $\{\mathbf{y}_i\}_i^{N_L}$.\\
 &  &  \textbf{Step 2}: For each generated lattice, find K-nearest neighbors in test dataset by KNN algorithm.\\
 &  &   \textbf{Step 3}: For each $i$-th generated lattice, Map $K$ neighbors to property space, obtaining $\{\mathbf{y}_{i,k}\}_k^{K}$.\\
 &  &   \textbf{Step 4}: For each $i$-th condition and corresponding $K$ properties $\{\mathbf{y}_k\}_k^{K}$, compute minimum euclidean distance.\\
\cline{2-3}
 & \multirow[c]{2}{*}{Efficiency}
 & $\bullet$ \textbf{Mean Evaluation Time (MET)}: Mean generation time per sample. \\
 &  & $\bullet$ \textbf{Mean Training Time (MTT)}: Mean training time per batch. \\
\hline
\multirow[c]{8}{*}{\rotatebox{0}{Prediction}} 
 & \multirow[c]{5}{*}{Accuracy }
 &  $\bullet$ $\text{MAE} = \frac{1}{n}\sum_{i=1}^{n} \Vert\mathbf{y}_i - \hat{\mathbf{y}}_i\Vert_1$, where $\mathbf{y}$ and $\hat{\mathbf{y}}$ denote the predicted and ground truth properties.\\
 &  & $\bullet$ $\text{NRMSE} = \frac{\sqrt{\frac{1}{n}\sum_{i=1}^{n}\Vert\mathbf{y}_i - \hat{\mathbf{y}}_i\Vert^2}}{\max(\mathbf{y}) - \min(\mathbf{y})}.$\\
 &  & $\bullet$ $R^2 = 1 - \frac{\sum_{i=1}^{n}\Vert\mathbf{y}_i - \hat{\mathbf{y}}_i\Vert^2}{\sum_{i=1}^{n}\Vert\mathbf{y}_i - \bar{\mathbf{y}}\Vert^2},
    $
    where $\overline{\mathbf{y}}$ denotes the mean of the observed values.\\
\cline{2-3}
 & \multirow[c]{2}{*}{Efficiency} 
 &  $\bullet$ \textbf{Mean Evaluation Time (MET)}: Mean prediction time per batch.  \\
 &  &  $\bullet$ \textbf{Mean Training Time (MTT)}: Mean training time per batch. \\
\hline
FE Simulation
 & Stiffness Computation
 & Use high-fidelity Finite Element (FE) simulation for accurate mechanical properties calculations, incorporating simulation and visualization of asymptotic homogenization~\cite{ANDREASSEN2014488,ARABNEJAD2013249,homogenizationcode,OZDILEK2024103674} to evaluate physics consistency.
\\
\Xhline{2\arrayrulewidth}
\end{tabular}
\vspace{-1em}
\end{table*}
We develop a novel evaluation framework in the ML toolbox to evaluate ML models for metamaterial applications.
Adopting a multi-perspective approach, our evaluation framework is designed to provide robust and unbiased assessments of metamaterial models. This is accomplished by incorporating and adapting established metrics from previous works~\cite{CDVAE,uniftruss,Stiffness,Modulus,SyMat} and by developing new metrics that capture the unique characteristics of metamaterials. 
As illustrated in Table~\ref{tab:evaluation_toolbox}, property prediction task focuses on accuracy through a combination of three metrics, while generation task evaluates model performance based on the validity, diversity, and conditional effectiveness of the generated lattices. 
Additionally, the overall evaluation also considers the efficiency of both training and testing processes.  More details of the evaluation toolbox are illustrated in Appendix~\ref{app:metrics}.
Here, we provide several definitions for unbiased generative evaluation metrics.
First, we define a symmetric node as a node that can find its central symmetric counterpart within an error range: 
\begin{definition}[Symmetric Node]\label{def:symmetric_node}
        Consider node $i$ with coordinates $\mathbf{p}_i$, the node is a symmetric node if there exists another node $j$ in the structure that satisfies: 
        $\left\Vert \mathbf{p}_i + \mathbf{p}_j - 2\mathbf{p}_c\right\Vert_2 < \epsilon,$ where
        $\mathbf{p}_c$ denotes central coordinates in the structure, and $\epsilon$ is a positive hyperparameter.
\end{definition}
\noindent In addition, the symmetry degree of a node is defined as the error value of the corresponding "most symmetric" node pair divided by the distance between the central coordinates and the farthest node.
\begin{definition}[Symmetry Degree]\label{def:symm_degree}
         The symmetry degree of node $i$ in a structure is defined as:
           $ s_{degree_i} = \frac{\epsilon_{max} - s_{error_i}}{\epsilon_{max}},$
        where $\epsilon_{max} = \max_j{\Vert \mathbf{p}_c - \mathbf{p}_j \Vert_2}$, $s_{error_i} = \min_j{ \Vert \mathbf{p}_i + \mathbf{p}_j - 2\mathbf{p}_c\Vert_2}$, $\mathbf{p}_c$ denotes central coordinates in this structure, and $j$ is a node in this structure.
\end{definition}
\noindent We then introduce periodicity, denoted as $\mathcal{V}_P$, to assess the generated structures at the lattice level. This metric aims to evaluate whether the structures can repeat for constructing a lattice, 
Formally, we define the necessary condition of periodicity of a structure, \eg, if a lattice is periodically valid it must satisfy this definition, as follows.
\begin{definition}[Periodicity]\label{def:period}
    Given a structure with node positions $\mathbf{P}$ and lattice vectors $\mathbf{L}$, for each dimension $d \in \{0, 1, 2\}$, there exist at least one pair of coordinate points $\mathbf{p}_i$ and $\mathbf{p}_j$ s.t. $\| (\mathbf{p}_i + \mathbf{l}_d) - \mathbf{p}_j \|_1 < \epsilon$, where $\|\cdot\|_1$ is the L1 norm and $\epsilon$ is the tolerance range.

\end{definition}


In addition to ML-level evaluation, we also include an FE simulation tool in this toolbox, which can accurately predict elastic properties given the lattice graphs. This simulation tool enables researchers to make more informed decisions. 

In summary, we develop the evaluation toolbox from 5 perspectives (\ie, validity, diversity, conditional effectiveness, accuracy, and efficiency) with 12 metrics, including an FE simulation tool for physics-aware computation.
\vspace{-0.5em}
\subsection{\fname: Visual-Interactive Interface Development}\label{sec:interface_dev}
\begin{figure}[b]
    \centering
    \vspace{-1em}
    \includegraphics[width=\linewidth]{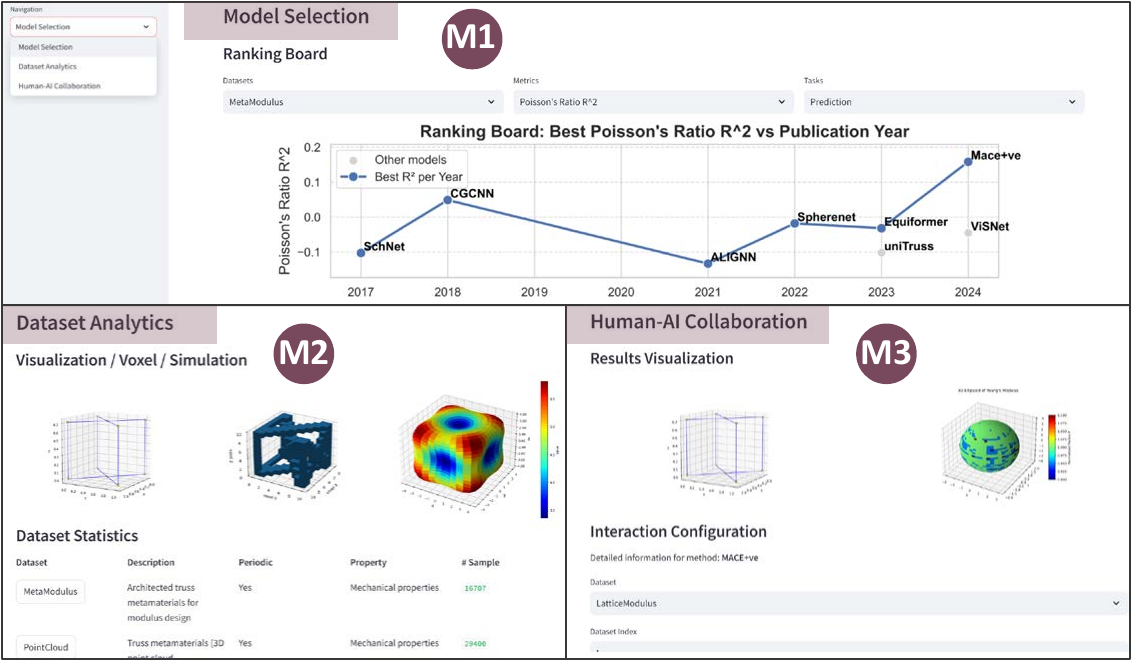}
    \vspace{-2.5em}
    \caption{Overview of the visual-interactive interface.}
    \vspace{-1em}
    \label{fig:interface}
\end{figure}

At the user level, \fname\ provides a web interface to address the human-AI collaboration challenge (C3). The web interface consists of three main modules as shown in Figure~\ref{fig:interface}, \ie, Ranking Board (M1), Dataset Interaction (M2), and Model Interaction (M3), that mitigate human-AI dual black-box issue.
\paragraph{M1: Model Selection} The Model Selection module provides a ranking board for overview of the advanced ML models' performance on metamaterials. Users can choose the datasets, tasks, and metrics they are interested in for visualization.
\paragraph{M2: Dataset Analytics} This module provides an interface for users to interact with data level, analyzing the intrinsic data distributions. Users can not only view overall dataset statistics but also visualize and simulate the metamaterials and their corresponding properties. This helps researchers identify the dataset they need.
\paragraph{M3: Human-AI Collaboration} This module enables easy calls for the proposed toolbox at the ML level. Users can specify the ML model and the datasets they wish to use, specify samples or conditions, and make predictions or generate results. The outcomes are visualized and can be simulated through the interface. By visualizing results and allowing parameter tweaks, the interface helps demystify the dual black-box of AI-assisted metamaterial design.
\begin{table}[tb]
    \vspace{-1em}
    \small
    \centering
    \caption{Data sanitization analysis.}
    \vspace{-1em}
    \begin{tabular}{c|cccc}
    \Xhline{2\arrayrulewidth}
        Dataset & $\mathcal{V}_{DR}$\% & $\mathcal{V}_{C}$\% &
        $\mathcal{V}_{S}$\% & $\mathcal{V}_{P}$\% \\
    \hline
        Original & 15.10 & 62.71 & 90.06 & 99.16 \\
        Processed & 24.07 & 100.00 & 94.92 & 99.13 \\
    \Xhline{2\arrayrulewidth}
    \end{tabular}
    \label{tab:sanit_modulus}
    \vspace{-1.5em}
\end{table}
\vspace{-0.5em}
\section{Results and Analysis}
In this section, we conduct extensive experiments to evaluate the efficacy of the proposed framework \fname at the data level, ML level, and user level. 

\vspace{-0.5em}
\subsection{Data Validity Analysis}\label{sec:analy_data}
Here we aim to show the effectiveness of the proposed unified data sanitization, thus mitigating C1.
Table~\ref{tab:sanit_modulus} shows the dataset statistics before and after our sanitization process (using MetaModulus as an example). We applied the validity evaluation metrics regarding four levels, \ie, dangling restrictions $\mathcal{V}_{DR}$, connectivity $\mathcal{V}_C$, symmetry $\mathcal{V}_S$ and periodicity $\mathcal{V}_P$ as in Table~\ref{tab:evaluation_toolbox}, to both the original and the sanitized versions of the dataset.
We observe that the validation metrics ($\mathcal{V}_{DR}$, $\mathcal{V}_{C}$, and $\mathcal{V}_{S}$) have increased after data sanitization, and $\mathcal{V}_{P}$ value maintains more than $99\%$. The results demonstrate the effectiveness of the unified sanitization process. More statistical details of the database are provided in Appendix~\ref{app:dataset}.

\vspace{-0.5em}
\subsection{Algorithms Comparison}
\begin{table*}[h!]
\small
\centering
\caption{Generation Evaluation. DR: Dangling Restriction, Conn: Connectivity, Sym: Symmetry, Peri: Periodicity, CR: Coverage Recall, CP: Coverage Precision, MET: Mean Evaluation Time (generation time per sample), MTT: Mean Training Time.}
\vspace{-1em}
\begin{tabular}{l|ccccc|ccc|c|rr}
\Xhline{2\arrayrulewidth}
\multirow{2}{*}{Approach} & \multicolumn{5}{c|}{Validity $\uparrow$} & \multicolumn{3}{c|}{Diversity $\uparrow$} & Cond. Effectiveness $\downarrow$ & \multicolumn{2}{c}{Efficiency} \\

                          & $\mathcal{V}_{DR}$\%   & $\mathcal{V}_{C}$\% & $\mathcal{V}_{S}$\%  & $\mathcal{V}_{P}$\% & Mean & Cov R.\%   & Cov P.\%   & Mean &  & MET (s)  & MTT (ms)    \\
\hline
\multicolumn{9}{l}{\textit{\textbf{Molecule targeted methods.}}}\\
EDM~\cite{EDM} &  N/A     &  N/A     & 0.00  &   0.00   &    0.00    &  0.00    &  0.00  &  0.00   &  982.13    &  3.18    & 161.39       \\
GeoLDM~\cite{GEOLDM} &  N/A    &   N/A   &  0.04    &  0.00    & 0.02     &   0.00   & 0.00  &   0.00   &  60.59    &    2.84  & 606.80   \\
\hline
\multicolumn{9}{l}{\textit{\textbf{Crystal targeted methods.}}}\\
CDVAE~\cite{CDVAE}   & N/A & N/A & 57.03  & 0.40    &  28.72   & 55.85   & 95.80  & 75.83 & N/A & 93.00  & 97.42    \\
DiffCSP ~\cite{diffcsp}   &   N/A   &   N/A   &  34.46   &   6.50   &   20.48     &   95.80   &   96.65   &  96.23  & N/A  &  2.97    &   63.79      \\
EquiCSP~\cite{EquiCSP}   &   N/A   &   N/A   &  55.37   &   3.55   &    29.46    &   100.00   &  52.35   &    76.18    &   N/A   &   1.90   &  64.57       \\
Cond-CDVAE~\cite{condCDVAE}   & N/A & N/A  &   19.37   &   2.00  & 10.69  & 68.60 &  80.50      & 74.55 & 0.2050 & 225.01 & 314.51    \\
SyMat~\cite{SyMat}   & N/A  & N/A  & 41.10     & 0.00     & 20.55      & 79.34     & 38.90     &  59.12     &   N/A   &  89.49    & 141.20        \\
CrystaLLM~\cite{CrystaLLM}   &   3.60   &   26.90   &  76.43   &   92.10   &  49.76      &   100.00   &   100.00   &  100.00      &   0.0983   &  2.08  &    14.51s     \\
Crystal-Text-LLM~\cite{gruver2024finetuned} & 23.50  &  68.50    &    89.37  &   96.10   &  69.37    &    100.00    &   100.00   & 100.00     &   0.0916     &    46.49   &       708.00        \\

\Xhline{2\arrayrulewidth}
\end{tabular}
\vspace{-1em}
\label{tab:generation}
\end{table*}

\begin{table*}[h!]
\small
\centering
\caption{Prediction Evaluation. MAE: Mean Absolute Error, R2: R-squared, NRMSE: Normalized Root Mean Square Error, MET: Mean Evaluation Time, MTT: Mean Training Time.} \label{tab:prediction}
\vspace{-1em}
\begin{tabular}{l|ccc|ccc|ccc|rr}
\Xhline{2\arrayrulewidth}
\multirow{3}{*}{Approach} & \multicolumn{9}{c|}{Accuracy} & \multicolumn{2}{c}{Efficiency (ms)} \\
                          & \multicolumn{3}{c|}{Young's Modulus} & \multicolumn{3}{c|}{Shear Modulus} & \multicolumn{3}{c|}{Poisson's Ratio} & \multicolumn{2}{c}{}\\
                          & MAE $\downarrow$  & R$^2$ $\uparrow$  & NRMSE $\downarrow$ & MAE $\downarrow$  & R$^2$  $\uparrow$  & NRMSE $\downarrow$ & MAE $\downarrow$ & R$^2$ $\uparrow$  & NRMSE $\downarrow$ & MET  & MTT \\
\hline
\multicolumn{9}{l}{\textit{\textbf{Molecule targeted methods.}}}\\
SchNet~\cite{schnet}     
    & 0.0005704    & 0.2304   &  0.1436   &0.0001343   & 0.01441   &   0.3958   & 0.3962   & -0.1025  & 0.02029   & 3.34 & 12.39\\
Spherenet~\cite{spherenet}     
    & 0.0004744 & 0.4548     & 0.1185 & 0.0001039  & 0.2839   & 0.08682 & 0.3561   & -0.01795 & 0.02499 &  11.62 & 33.08\\
Equiformer~\cite{equiformer}     
    & 0.0006669 & -0.3892    & 0.2469 & 0.0002226 & -1.1149   & 0.3459  & 0.3673   & -0.03171 & 0.05805 & 63.17   & 204.70 \\
ViSNet~\cite{visnet}  
    & 0.0006223 & 0.05871    & 0.1506 & 0.06375   & 0.06375  & 0.1003  & 0.3699  & -0.04497 & 0.01916 & 15.63   & 32.33 \\
\hline
\multicolumn{9}{l}{\textit{\textbf{Crystal targeted methods.}}}\\
CGCNN~\cite{CGCNN}    
    & 0.0006179     & 0.3550     &   0.1785    & 0.0001475     & 0.09353     & 0.1720     &   0.3922    & 0.04905     & 0.08282     & 10.76     & 48.51\\
ALIGNN~\cite{ALIGNN}    
    &   0.0008320   &   -1.2955   &   0.1544    &   0.0001460   &   -0.01672   &   0.09749   &   0.31267    &   -0.13298   &  0.01634  & 990.34  &   44993.19  \\
\hline
\multicolumn{9}{l}{\textit{\textbf{Metamaterial targeted methods.}}}\\
uniTruss ~\cite{uniTruss}   
    & 0.0006266 & 0.1812  & 0.1389  & 0.0001451 & 0.06374  & 0.09955  & 0.3970   & -0.1016  & 0.02014 & 1.18   & 0.22 \\
Mace+ve~\cite{mace_ve}   
    & 0.0003882 & 0.6692  & 0.08797 & 0.0001211 & 0.1932   & 0.08913  & 0.2881   & 0.1585   & 0.01738 & 27.34   & 304.2 \\
\Xhline{2\arrayrulewidth}
\end{tabular}
\vspace{-1em}
\end{table*}

Below we explore how the complex ML models perform on metamaterial applications. We employ our evaluation metrics to compare the integrated algorithms shown in Table~\ref{tab:comparisonModel} regarding both generative and predictive tasks.
We train all models on A100 GPUs following each baseline’s original hyperparameters and training strategies. We primarily use the MetaModulus dataset (16,707 samples, split 8,000/2,000/6,707 for train/valid/test) with its three mechanical properties (Young’s modulus, Shear modulus, Poisson's ratio). Our baselines target molecules, crystals, or metamaterials as shown in Table~\ref{tab:comparisonModel}. Methods that cannot handle lattices or edges are adapted accordingly; Large Language Models (LLMs) are pre-trained on crystals and fine-tuned on metamaterials. More implementation details are stated in Appendix~\ref{app:imple}.
To comprehensively evaluate these models, we utilize the proposed evaluation toolbox in Table~\ref{tab:evaluation_toolbox} for evaluation. Additional details appear in the Appendix.
\paragraph{Benchmarking Generative Models} Table~\ref{tab:generation} compares the generation performance of generative models through the evaluation toolbox. In general, we have the following observations:
     \emph{(1) Periodicity constraints benefit generation}: EDM and GeoLDM (top of Table~\ref{tab:generation}) are molecule-targeted methods. Hence, they do not satisfy crystal-specific constraints (\eg, periodicity) as shown in Table~\ref{tab:comparisonModel}. Table~\ref{tab:generation} suggests that they cannot generate valid and diverse structures. Specifically, it shows up as 0 (or near 0) ``validity'' on metrics tied to periodicity ($\mathcal{V}_p$) and symmetry ($\mathcal{V}_s$), and also 0 coverage (Cov. R and Cov. P) of test data space. By contrast, crystal targeted methods (\eg, CDVAE, DiffCSP, EquiCSP, \etc.) that generally conduct periodicity constraints show higher symmetry and periodicity validities ($\mathcal{V}_s$ and $\mathcal{V}_p$). We suspect that the molecular-based methods without periodicity constraints cannot adapt to metamaterial generation.
     \emph{(2) Equivariance tends to boost validity and diversity}: Comparing mean validity of equivariant architectures (EquiCSP and DIffCSP), semi-equivariant architectures (CDVAE and Cond-CDVAE), and invariant architectures (SyMat), the performance decreases accordingly from 29.48\% and 20.48\% (equivariant), 10.69\% and 28.72\% (semi-equivariant), to 10.64\% (invariant) as per validity; and 96.23\% and 76.18\% (equivariant), 75.83\% and 74.55\% (semi-equivariant) to 0.00 (invariant) as per diversity.
     \emph{(3) LLM approaches excel in all metrics without geometric constraints}: CrystaLLM and Crystal-Text-LLM do not declare explicit geometric constraints (regarding periodicity and symmetry). However, they show superior performance on validity, which may be because the LLMs, despite lacking explicit geometric constraints, can learn valid periodicity and symmetry from a large number of data (both pre-train and finetune data). In addition, their 100\% coverage on the test dataset demonstrates their larger design space compared to other methods with geometric constraints. Moreover, their low Conditional Effectiveness indicates adherence to desired conditions during generation.
     \emph{(4) LLM approaches are training inefficient. More constraints lead to longer generation time.} Focusing on mean training time per batch (MTT), it can be concluded that the training time of LLM-based methods is many times longer than others. In addition, regarding the generation time per sample (MET), methods with more geometric constraints tend to be less generation efficient.
     
\paragraph{Benchmarking Predictive Models} We benchmark the predictive models in Table~\ref{tab:prediction}, from which we can have the following observations.
     \emph{(1) Metamaterial oriented methods perform better}: Comparing the three design targets, the metamaterial targeted methods generally perform superior to the other, especially Mace+ve outperforms the second best 0.2144 and 0.1765 regarding R$^2$ on Young's modulus and Poisson's ratio, respectively. uniTruss obtains the best efficiency, although its accuracy is moderate.
    This superior performance of metamaterial-tailored methods is reasonable since they align well with the dataset.
    \emph{(2) Periodic constraints are ineffective for accuracy}: Comparing the methods with periodicity constraints (\ie, ViSNet, CGCNN, and ALIGNN), they do not have obvious superiority to the methods without periodicity constraints (\eg, SchNet, SphereNet, Mace+ve, and uniTruss). This observation is different from generation task, and we suspect it is because the mechanical properties is not related to periodicity.
    \emph{(3) Invariant models are efficient and effective}: Comparing equivariant models (\ie, Equiformer, ViSNet, and Mace+ve) and invariant models (other models), the invariant models demonstrate greater efficiency on both the evaluating and the training phase. Moreover, most invariant models perform better than equivariant models except Mace+ve which is specifically designed for the conducted dataset.
    
In summary, metamaterial-specific methods (\eg, Mace+ve) perform the best for mechanical property prediction, and LLM-based method is most effective for metamaterial generation. This benchmarking provides guidance on selecting appropriate models for metamaterial research (addressing C2).

\vspace{-0.5em}
\subsection{Case Study on Visual-Interactive Interface}
\begin{figure}
    \centering
    \includegraphics[width=\linewidth]{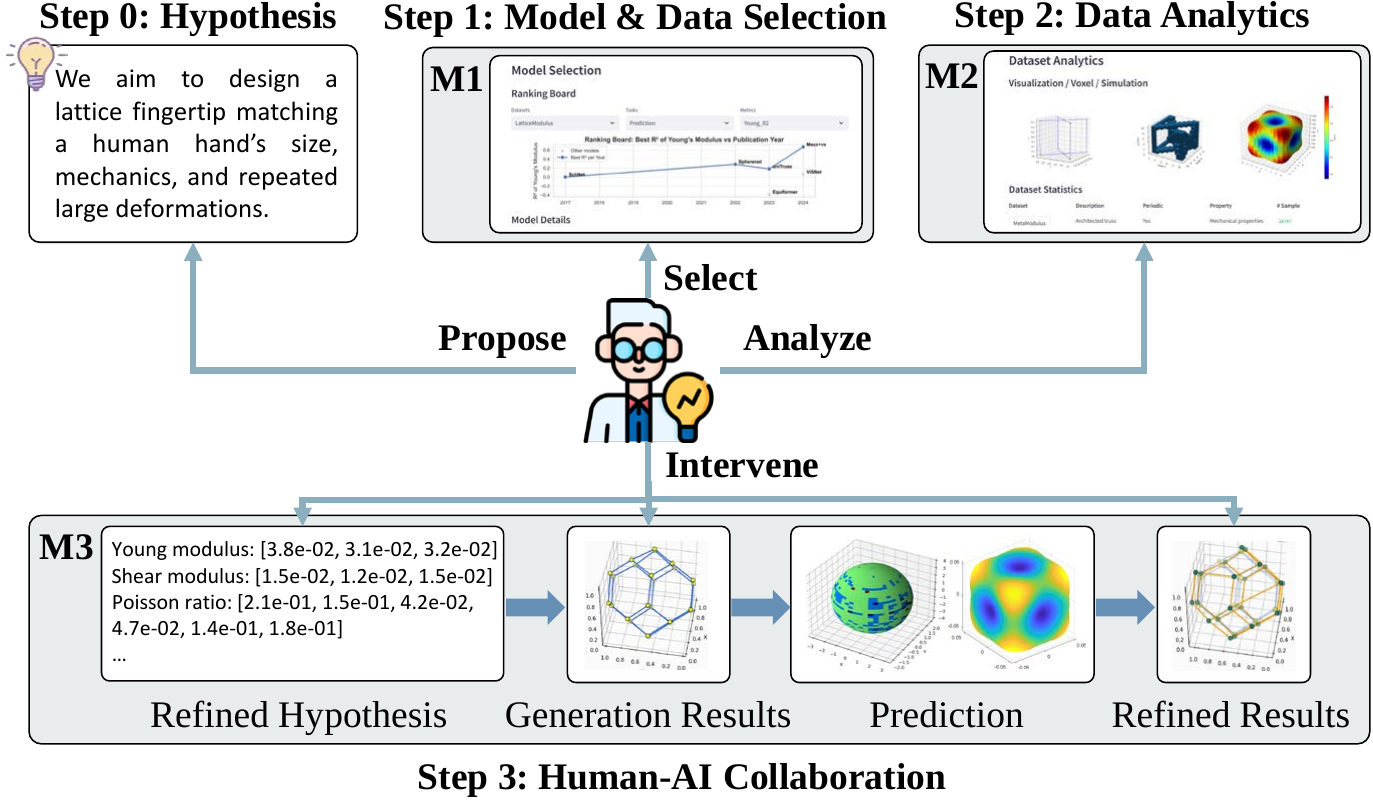}
    \vspace{-2em}
    \caption{A case study on human-AI collaboration for metamaterial discovery.}
    \vspace{-2em}
    \label{fig:case_study}
\end{figure}

Here we provide a case study on how \fname\ enhances metamaterial design for a specific hypothesis.
By integrating the three key modules, the system guides metamaterial researchers from model selection through dataset analytics to predictive simulation, ultimately accelerating the discovery of effective metamaterials.
To be specific, in this case study shown in Figure~\ref{fig:case_study}, the goal is to design a lattice structure for the fingertip with desired mechanical properties—such as Young’s modulus, Shear modulus, Poisson's ratio, achieving a balance between stiffness and flexibility. 
The process begins with \textbf{Step 0 (Hypothesis)}, where researchers define the requirement for a fingertip structure that replicates a real hand’s size and mechanics. Moving to \textbf{Step 1 (Model and Data Selection)}, they leverage M1 in the web interface to identify suitable ML models and datasets. In \textbf{Step 2 (Data Analytics)}, the chosen datasets are analyzed to obtain insights into structural-performance relationships, guiding the refinement of the lattice design. Finally, \textbf{Step 3 (Human-AI Collaboration)} is where M3 facilitates iterative design: researchers propose modifications based on domain knowledge according to their analysis of Step 2, while the AI predicts the fingertip’s mechanical responses and generate specific lattice structures, leading to rapid refinements of both the hypothesis and the metamaterial structure. As a result, users found the visualizations of refined results intuitive.
This loop of expert feedback and AI-driven prediction expedites the development of a lattice fingertip optimized for finger-like mechanical performance. 
\vspace{-1.em}
\section{Conclusion and Future Work}
\vspace{-0.3em}
In this paper, we propose a multi-level system, \fname, to bridge the gap between traditional metamaterial research and advanced ML methodologies. At the data level, our unified data processing and representation framework addresses the inherent data heterogeneity in metamaterial datasets. The intermediate ML level handles model complexity by offering an extensive ML toolbox—comprising both a model suite and a multi-perspective evaluation toolkit—while the top level addresses dual black-box challenge between human and the AI system through a visual-interactive web interface that reduces the opacity of human-AI collaboration. Our experimental results demonstrate enhanced data validity, provide an in-depth analysis of ML model performance in metamaterial contexts, and illustrate the system's potential to accelerate metamaterial discoveries with a case study.

Looking forward, we propose two research directions to advance both ML and metamaterials. \textbf{Q1}: Design ML models that integrate geometric constraints unique to metamaterials. \textbf{Q2}: Strengthen collaborations between metamaterial researchers and AI systems to drive innovative breakthroughs.




\bibliographystyle{ACM-Reference-Format}
\bibliography{main}

\clearpage
\appendix

\section{Related Works}\label{app:relatedwork}
\subsection{Material Benchmarks}
There are some existing works on the material benchmark, for example, MatBench~\cite{MatBench}, Geom3D~\cite{Gom3D}, OMat24~\cite{omat24}, and M$^2$Hub~\cite{M2Hub}. To be specific, MatBench~\cite{MatBench} focuses on material property prediction, including mechanical properties, elastic properties, electronic properties, optical and phonon properties, and thermodynamic stabilities. Geom3D~\cite{Gom3D} organizes a benchmark on property prediction for small molecules, proteins, and crystalline materials. The predicted properties include quantum properties, molecular dynamics, energy, etc. Besides property prediction, Geom3D~\cite{Gom3D} also proposes a pipeline for material geometric pretraining, i.e., pretraining solely on the spatial arrangement of atoms. OMat24~\cite{omat24} focuses on predicting the energy, forces, and cell stress for crystalline materials. Since OMat24~\cite{omat24} aims at finding new material structures, the majority of the structures in its datasets are in a non-equilibrium state, \ie, not able to exist stably owing to lack of symmetry, periodicity, \etc. M$^2$Hub~\cite{M2Hub} provides a benchmark on both material property prediction and generation. The concerned material types include crystalline materials, molecules, bulks and some other non-periodic materials. In the aspect of material generation, M$^2$Hub~\cite{M2Hub} evaluates reconstruction precision, the validity of generated materials, and the distribution of generated materials. 

However, none of the benchmarks are for the metamaterial domain. OMat24~\cite{omat24} focuses on non-equilibrium structures, which is still invalid for metamaterials. MatBench~\cite{MatBench}, M$^2$Hub~\cite{M2Hub}, and Geom3D~\cite{Gom3D} all involve periodic materials, but such structures are all naturally existing ones, which may have different chemical formulas, but very similar or even identical geometric arrangement. The structures of metamaterial, however, do not have to follow the natural structures so long as they satisfy some basic requirements like periodicity. Therefore, the periodic material structures provided in existing benchmark works only occupy a very limited part of the design space of metamaterial, although they are possible for metamaterial design. Moreover, the existing evaluation tools do not fully satisfy the requirements of the metamaterial benchmark. Owing to the considerable discrepancy between metamaterial and conventional material structures, the evaluation of metamaterial generation tasks needs to be largely different from that of conventional materials.


\subsection{Geometric ML for Molecule, Crystal, and Metamaterial}
Works of \cite{PotentialNet,schnet,spherenet,gasteiger2021gemnet,EquiformerV2,satorras2021n,atz2021geometric,MACE,visnet} propose various geometric ML methods for molecular property prediction, molecule generation, and molecular dynamics simulations, while \cite{CDVAE,SyMat,diffcsp,CGCNN,zeni2025generative} focus on geometric ML-based crystal materials for material generation and chemical property prediction. More recently, \cite{mace_ve} applied a specific geometric ML method~\cite{MACE} for stiffness prediction, but it targets only a specific model and mechanical property, leading to a loss of generalizability.

Geometric machine learning (ML) has been widely explored to model 3D-structured atomic graphs, \eg, molecules~\cite{PotentialNet,schnet,spherenet,gasteiger2021gemnet,EquiformerV2,satorras2021n,atz2021geometric,MACE,visnet} and crystals~\cite{CDVAE,SyMat,diffcsp,CGCNN,zeni2025generative}. These studies incorporate abundant 3D structural information, such as translation, rotation, and inversion symmetries in Euclidean space, invariance and equivariance in models, and periodic boundary conditions along three dimensions. Despite the tremendous progress, these methods primarily focus on atomic graphs with chemical properties.

\section{Model Toolbox Details}\label{app:ml_models}
We provide a use case of the model toolbox for training predictive models and tests. All aggregated models can be easily trained and tested by the following five-steps codes.
\begin{verbatim}
# Step 0: Edit the config 
   file at: configs/[Model]/[Dataset]_config.yml
# Step 1: Initialize a model toolbox Object with
    configs by indicating [model_name] and [dataset].
model = MaceVeModel(model_name='mace_ve', 
              dataset_name='LatticeModulus', 
              device=device, root_path='./') 
# Step 2: Load dataset.
model.load_data()
# Step 3: Load model.
model.load_model()
# Step 4: Training model.
model.train()
# Step 5: Testing model.
r2, nrmse, mae = model.test()
\end{verbatim}
    
    


\section{Evaluation Toolbox Details}\label{app:metrics}
\begin{figure}[b]
    \centering
        \includegraphics[width=\linewidth]{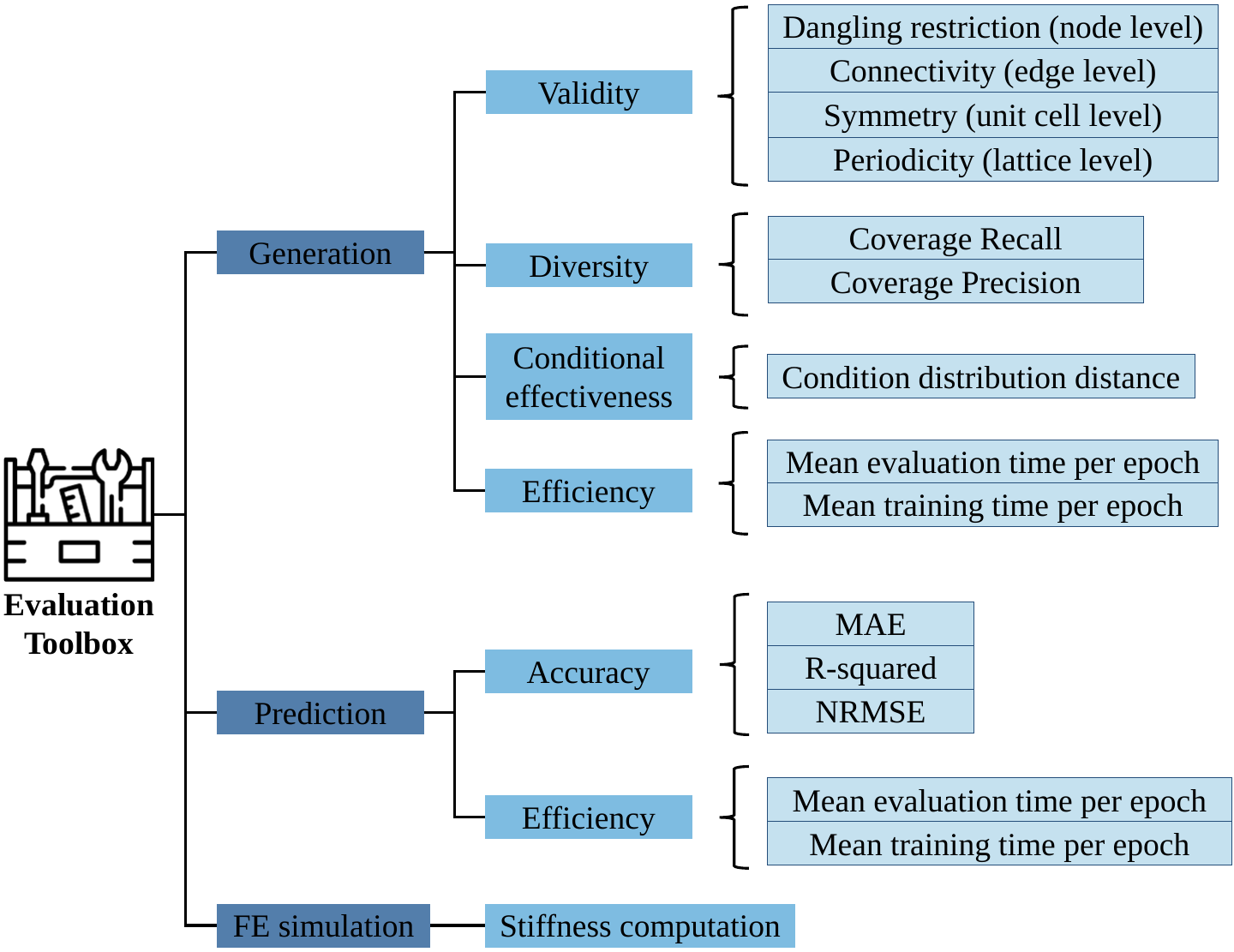}
    \caption{Overall framework of proposed evaluation toolbox.}
    \label{fig:evaluationtoolbox}
\end{figure}
The overall framework of the proposed evaluation toolbox is shown in Figure~\ref{fig:evaluationtoolbox}. Specifically, it contains generative task and predictive task evaluations.
\paragraph{Generative Task Evaluation} We propose to evaluate the generation performance of lattice generation models from four perspectives, \ie, validity, diversity, and conditional effectiveness ratio and efficiency.

\textbf{Validity}. Considering $N_L$ lattice structures are generated, we evaluate the validity of generated lattices from four levels.
\textit{(1)~Dangling restriction (node level)}. The generated structure is considered invalid at node level only if it contains one dangling node, \ie, a node holds less than one edge. The dangling restriction ratio ($\mathcal{V}_{DR}$) is computed as:
    $
        \mathcal{V}_{DR} = 1 - \frac{N_{D}}{N_L},
    $
    where $N_{D}$ is the number of structures that contain more than one dangling node.
\textit{(2)~Connectivity (edge level)}. The generated structure is invalid at the edge level only if the structure is not a connective graph. We compute the ratio of the generated connected graphs:
    $
        \mathcal{V}_{C} = \frac{N_{C}}{N_L},
    $
    where $N_{C}$ is the number of structures that are connected to the graph.
\textit{(3)~Symmetry (unit cell level)}. We propose to evaluate the symmetry of a structure by computing the central symmetry ratio ($\mathcal{V}_{S}$) of a graph in the 3D Cartesian space. Specifically, $\mathcal{V}_{S}$ is defined as:
\begin{equation}
    \mathcal{V}_{S} = \frac{1}{N_L}\sum_{k}^{N_L}\frac{N_{S_k}\cdot\sum_i^{N_k}s_{degree_i}}{{N_k}^2},
\end{equation}
where $N_L$ is the number of generated structures, $N_k$ is the node number of $k$-th structure, and $N_{S_k}$ is the number of Symmetrical Node that is defined in Definition.~\ref{def:symmetric_node} in $k$-th structure, and $s_{degree_i}$ denotes Symmetry Degree that is defined in Definition~\ref{def:symm_degree}.
In detail, we define a symmetrical node as a node that can find central symmetrical ones within an error range: 
\begin{definition}[Symmetrical Node]
        $\mathbf{p}_c$ denotes central coordinates in this structure, and $\epsilon$ is a positive hyperparameter. We consider node $i$ with coordinates $\mathbf{p}_i$ to be a symmetrical node iff exists another node $j$ in the structure satisfies: 
        $
            \left\Vert \mathbf{p}_i + \mathbf{p}_j - 2\mathbf{p}_c\right\Vert_2 < \epsilon.
        $
\end{definition}
\noindent In addition, the symmetry degree of a node is defined as the error value of the corresponding "most symmetric" node pair divided by the distance between the central coordinates and the farthest node.
    \begin{definition}[Symmetry Degree]
        $\mathbf{p}_c$ denotes central coordinates in this structure, and $j$ is a node in this structure. The symmetry degree of node $i$ in a structure is defined as:
           $ s_{degree_i} = \frac{\epsilon_{max} - s_{error_i}}{\epsilon_{max}},$
        where $\epsilon_{max} = \max_j{\Vert \mathbf{p}_c - \mathbf{p}_j \Vert_2}$, and $s_{error_i} = \min_j{ \Vert \mathbf{p}_i + \mathbf{p}_j - 2\mathbf{p}_c\Vert_2}$.
    \end{definition}
\noindent \textit{(4)~Periodicity (lattice level)}. According to Definition~\ref{def:lattice}, a lattice is formed by periodically repeating unit cell structures along the lattice vectors $\mathbf{L}$. Therefore, we introduce periodicity, denoted as $\mathcal{V}_P$, to assess the generated structures at the lattice level. This metric aims to evaluate whether the structures can repeat for constructing a lattice, 
Formally, we define the necessary condition of periodicity of a structure:
    \begin{definition}[Periodicity]
    \label{def:lattice}
        Given a structure with node positions $\mathbf{P}$ and lattice vectors $\mathbf{L}$, for each dimension $d \in \{0, 1, 2\}$, there exist at least one pair of coordinate points $\mathbf{p}_i$ and $\mathbf{p}_j$ such that $\mathbf{p}_i + \mathbf{l}_d$ is approximately equal to $\mathbf{p}_j$ in the L1 norm within a tolerance range $\epsilon$. Formally,
        \begin{equation*}
            \begin{aligned}
            &\forall d \in \{0, 1, 2\},\\
            &\exists i \in \{0, 1, \ldots, N-1\}, \exists j \in \{0, 1, \ldots, N-1\},\\
            &\text{s.t. } \| (\mathbf{c}_i + \mathbf{l}_d) - \mathbf{c}_j \|_1 < \epsilon.
            \end{aligned}
        \end{equation*}
    \end{definition}
\noindent Eventually, the evaluation of the periodicity of generated lattices can be computed by $\mathcal{V}_P = \frac{N_P}{N_L}$, where $N_P$ denotes the number of generated structures that satisfy Definition~\ref{def:lattice}.

\textbf{Diversity}. To evaluate the diversity of generated lattices, we revise the diversity metrics~\cite{CDVAE,xu2021learning,GeoMol} that were originally proposed for atomic materials to apply to lattice structures. The motivation for the diversity evaluation is to utilize the test dataset and compute the overlap between the test dataset and generated structures. The pairwise distance between two structures with node coordinates is defined in previous works~\cite{CDVAE,zimmermann2020local} as $D(\mathbf{P}_i, \mathbf{P}_j)$, where $\mathbf{P}_i$ and $\mathbf{P}_j$ are the node positions of $i$-th structure and $j$-th structure in dataset respectively. Given $N_L$ generated structures with $\{\mathbf{P}_i\}_{i=1}^{N_L}$ and $N_t$ test structures with $\{\mathbf{P}^*_j\}_{j=1}^{N_t}$, the two diversity metrics are defined as follows. (1) \textit{Coverage recall}. Intuitively, coverage recall measures how many structures in the ground truth dataset are covered by generated structures,\ie,
    \begin{equation}
    \begin{aligned}
    \text{COV}_R=\frac{1}{N_t}\vert \{&i \in [1,\ldots,N_t]: \exists k \in [1,\ldots,N_L],\\
    &D(\mathbf{P}^*_i, \mathbf{P}_k) < \epsilon_{cov} \} \vert.
    \end{aligned}
    \end{equation}
(2) \textit{Coverage precision}. Similarly, coverage precision measures how many generated structures can find similar structures in the test dataset:
    \begin{equation}
    \begin{aligned}
    \text{COV}_P=\frac{1}{N_L}\vert \{&i \in [1,\ldots,N_L]: \exists k \in [1,\ldots,N_t],\\
    &D(\mathbf{P}_i, \mathbf{P}^*_k) < \epsilon_{cov} \} \vert.
    \end{aligned}
    \end{equation}

\begin{figure}[b]
    \centering
    \includegraphics[width=\linewidth]{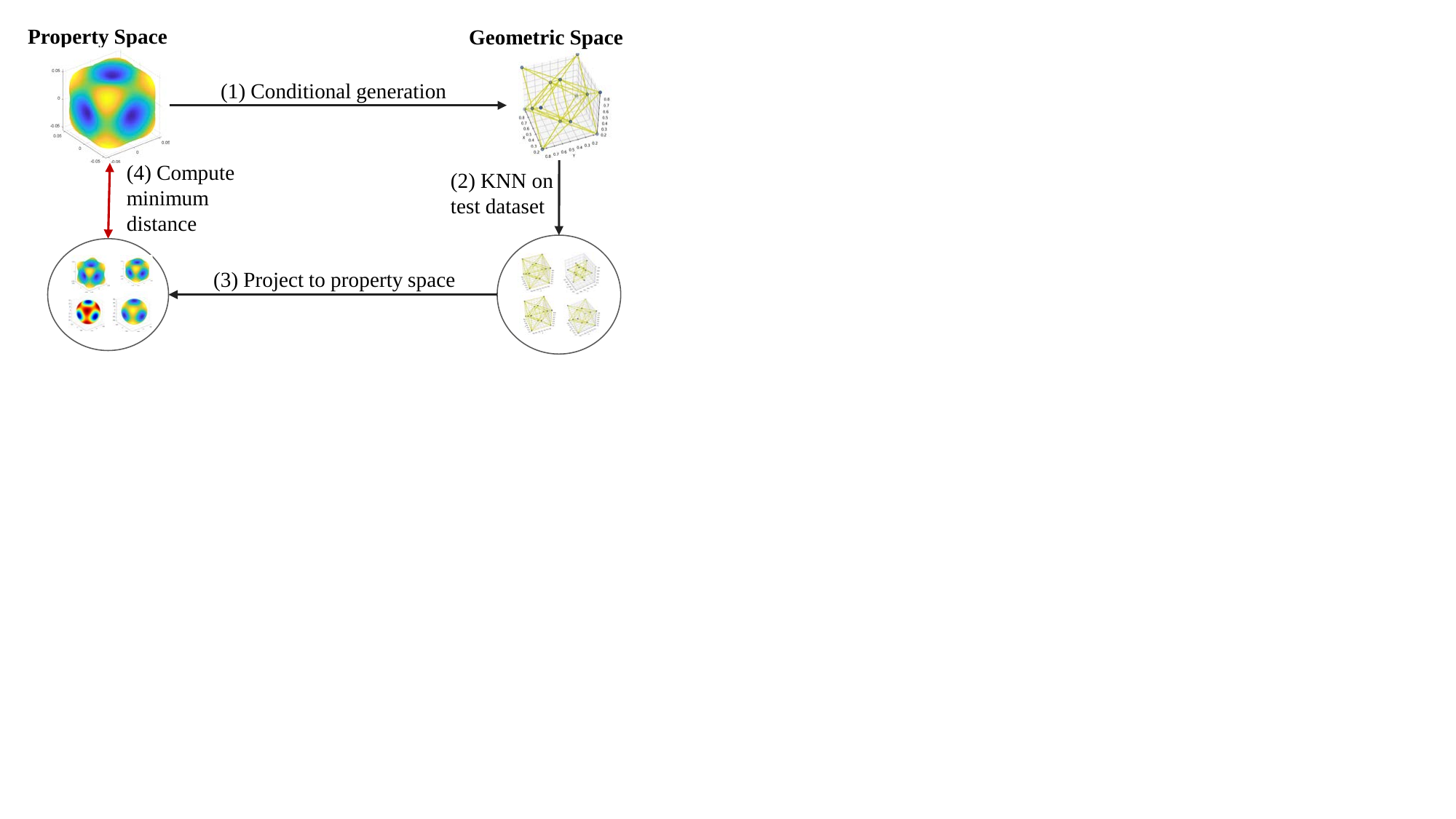}
    \caption{Computation process for assessing Condition Effectiveness. (1) After generating the lattices, (2) the K-nearest neighbor (KNN) algorithm is applied to the test dataset to identify various similar samples. (3) These samples are then projected to property space. (4) In the property space, the minimum Euclidean distance, representing the condition effectiveness, is calculated.}
    \label{fig:cond_eff}
\end{figure}
\textbf{Conditional Effectiveness}.
 In addition to the multi-level evaluation of validity and diversity, we propose a metric called \textit{conditional effectiveness} to assess how well a condition performs for conditional generation models. This statistic-based metric does not require complicated, time-consuming simulations to recompute mechanical properties, enabling fast approximations of condition effectiveness.
The intuitions behind the condition effectiveness include: (1) FE simulation is time-consuming, so we find the most similar structures in the dataset as an approximation. (2) Similar properties can correspond to various structures, while similar structures should share similar properties, so the most similar structure should be found in geometric space instead of property space. (3) Using One structure to represent the generated space may include bias. Therefore, we use a cluster of similar structures to approximate the generated lattice. Based on these intuitions, we sample a cluster of structures from the test dataset to approximate the generated structure and calculate the mean Euclidean distance between the cluster and the conditioned property. The process is shown in Figure~\ref{fig:cond_eff}.

\paragraph{Predictive Evaluation} Using $\mathbf{y}$ and $\hat{\mathbf{y}}$ denote the predicted and ground truth properties, the formulas for metrics MAE, NRMSE, and R$^2$ are defined as follows. (1) \textit{MAE} measures the average magnitude of errors in a set of predictions without considering their direction. It is calculated as:
    $$
    \text{MAE} = \frac{1}{n}\sum_{i=1}^{n} |y_i - \hat{y}_i|.
    $$
(2) \textit{NRMSE} is integrated into the evaluation toolbox due to the large variance among different property values. NRMSE is a normalized form of the Root Mean Square Error (RMSE), which can be used to compare the performance of models across datasets with different scales. The formula for NRMSE is:
    $$
    \text{NRMSE} = \frac{\sqrt{\frac{1}{n}\sum_{i=1}^{n}(y_i - \hat{y}_i)^2}}{\max(y) - \min(y)}.
    $$
(3) \textit{R$^2$} indicates the proportion of the variance in the dependent variable that is predictable from the independent variables. It ranges from negative infinity to 1, with higher values indicating a better fit of the model. The formula for R$2$ is:
    $$
    R^2 = 1 - \frac{\sum_{i=1}^{n}(y_i - \hat{y}_i)^2}{\sum_{i=1}^{n}(y_i - \bar{y})^2},
    $$
    where $\overline{y}$ denotes the mean of the observed values.

\section{Implementation Details}\label{app:imple}
\subsection{Experiment Settings} We run all models on A100 80G/40G GPU cards. We follow the hyperparameters (such as batch size, epochs, learning rate, \etc.) and training strategies (\eg, Exponential Moving Average, \etc.) implemented in their original codes to run all experiments.
\subsection{Dataset} In this section, we focus on MetaModulus since it contains the most abundant mechanical properties (\ie, Young's modulus, Shear modulus, and Poisson's ratio) compared to other datasets included in \fname, which can provide a more comprehensive comparison. This dataset comprises 16,707 samples in all, and we randomly split this dataset to 8000/2000/6707 for training/validation/testing for all models.

\subsection{Baselines and Metrics} We employ ML baselines shown in Table~\ref{tab:comparisonModel}, covering different geometric constraints (\eg, periodicity and symmetry), various backbones (\eg, Diffussion, VAE, LLM, \etc.), and three design targets (\eg, Molecule, Crystal, and Metamaterial). In detail, for molecular targeted methods that do not accept lattice representation \(\mathbf{L}\), we exclude this representation dimension during training and evaluation. For methods that do not accept edges \(E\) in the graph \(\mathcal{G}\), we modify their original code to ensure they can still capture the structural information. For LLM approaches, CrystaLLM is trained from scratch on an augmented metamaterial dataset with shuffled node orders, using the larger model version from the original paper. Crystal-Text-LLM is fine-tuned with LoRA on LLaMA-2-7B using the metamaterial dataset.
\section{More Results}\label{app:more_results}
\subsection{More Dataset Sanitization Results}\label{app:dataset}
\begin{figure}[t]
    \centering
    \includegraphics[width=\linewidth]{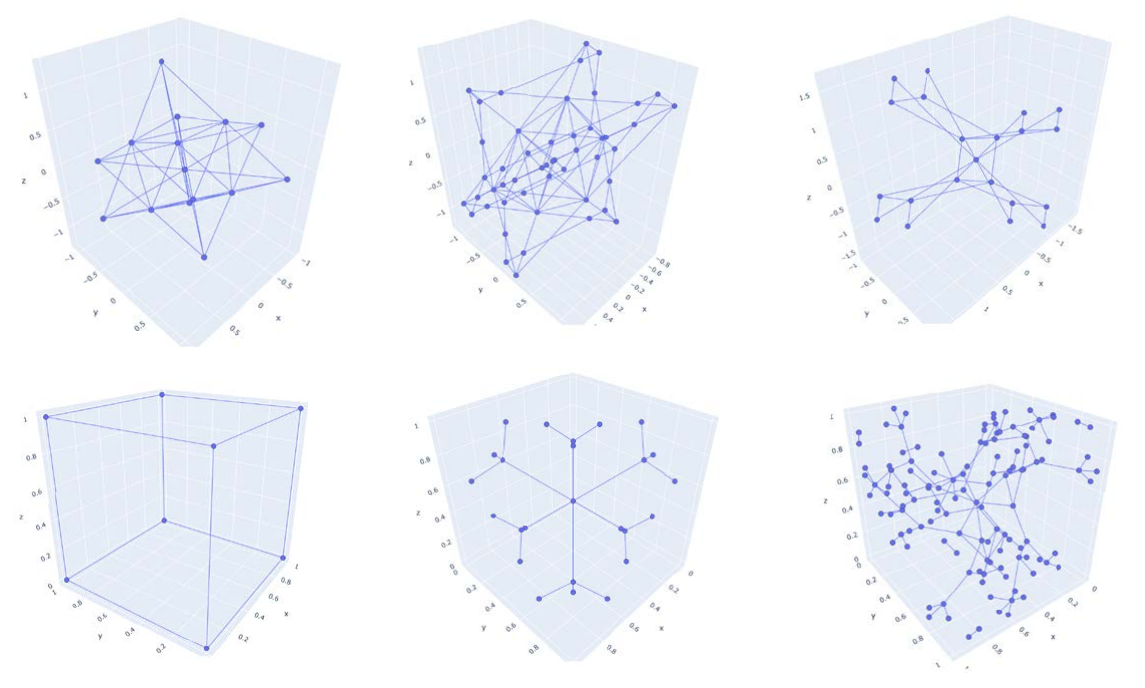}
    \caption{Visualization of 3D graph dataset.}
    \label{fig:data_3D_graph}
\end{figure}

\begin{figure}[t]
    \centering
    \includegraphics[width=\linewidth]{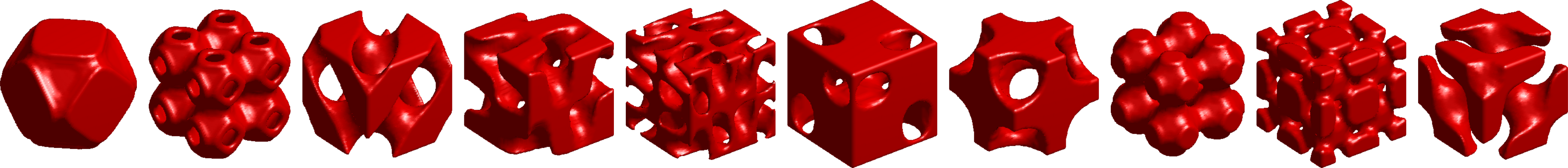}
    \caption{Visualization from Pointclouds dataset~\cite{chan2021metaset}.}
    \label{fig:data_cloud_point}
\end{figure}

Figures~\ref{fig:data_3D_graph} and \ref{fig:data_cloud_point} visualize several samples of MetaStiffness, MetaModulus, and PointCloud datasets.
\begin{figure}[t]
    \centering
    {
        \includegraphics[width=0.48\linewidth]{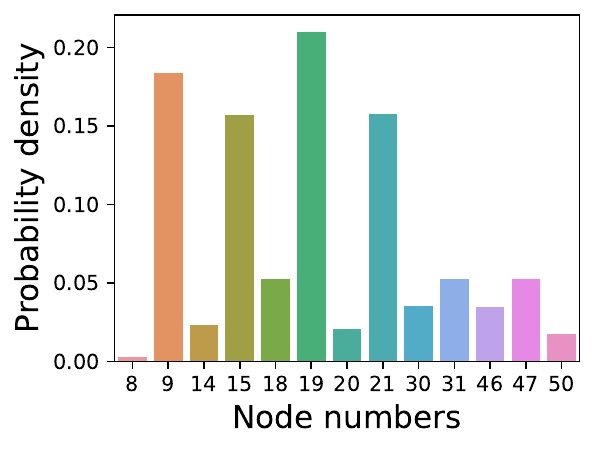}}
    {
        \includegraphics[width=0.48\linewidth]{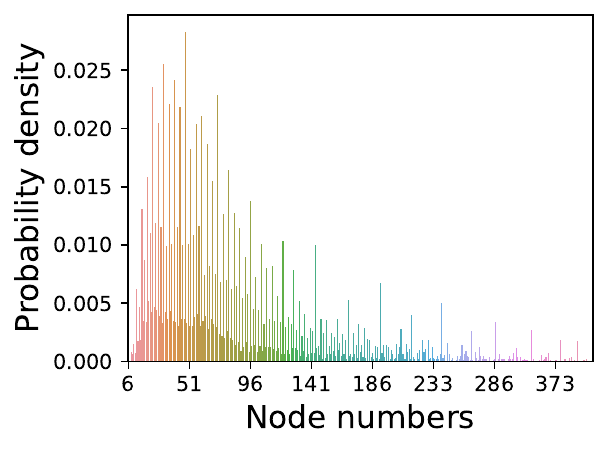}}
    \caption{Statistics of node numbers on MetaStiffness (left) and MetaModulus (right) dataset}
    \label{fig:data_statistics}
\end{figure}

In addition, we summarize the node number distribution of sanitized datasets in Figure~\ref{fig:data_statistics}. 
\begin{table}[tb]
\small
    \centering
    \caption{Data sanitization effectiveness on MetaModulus. $\mathcal{V}_{DR}$, $\mathcal{V}_{C}$, $\mathcal{V}_{S}$, and $\mathcal{V}_{P}$ indicates four-level data validity; Gini coef. implies node distribution heterogeneity. It shows that the sanitization increases the data validity and reduces the data heterogeneous.}
    \begin{tabular}{c|ccccc}
    \toprule
        Dataset &  $\mathcal{V}_{DR}$\%   & $\mathcal{V}_{C}$\% & 
        $\mathcal{V}_{S}$\%  & $\mathcal{V}_{P}$\% & Gini coef. of \# nodes\\
    \midrule
        Original & 15.10 & 62.71  & 90.06 & 99.16 &  0.423\\
        Processed & 24.07 & 100.00  & 94.92  & 99.13 &  0.394\\
    \bottomrule
    \end{tabular}
    \label{tab:app_sanit_modulus}
\end{table}
\begin{figure*}[t]
    \centering
    \subfloat[CCDF of original dataset.]{
        \includegraphics[width=0.35\linewidth]{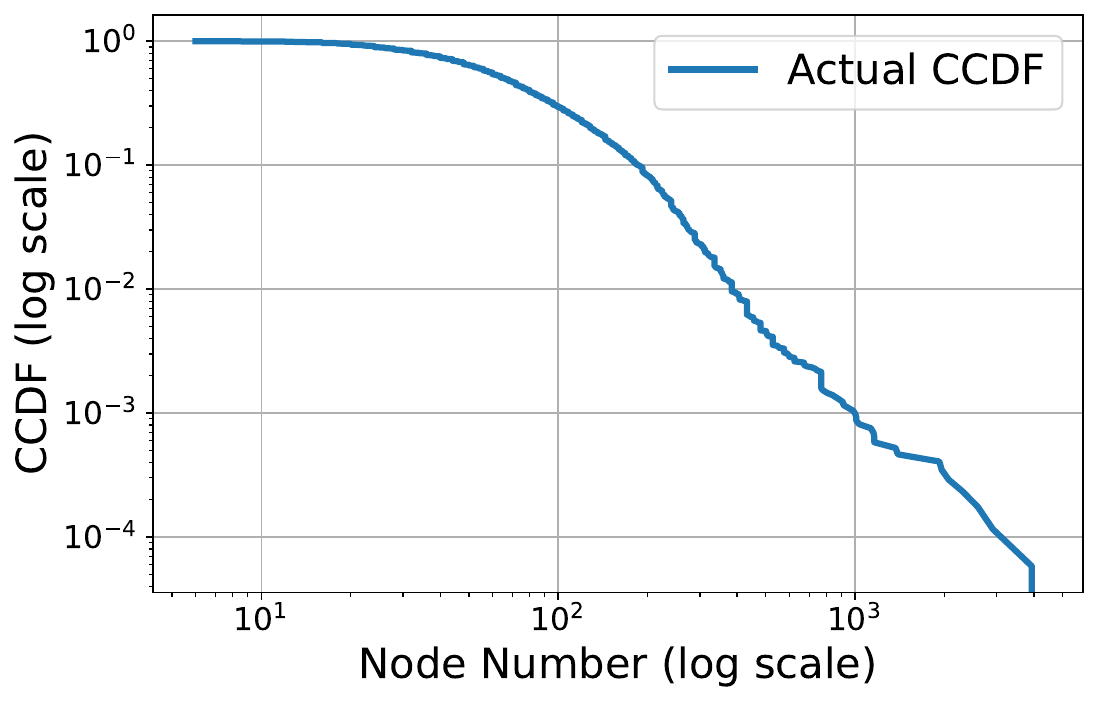}
        \label{}
        }
    \hspace{5em}
    \subfloat[CCDF of processed dataset.]{
        \includegraphics[width=0.35\linewidth]{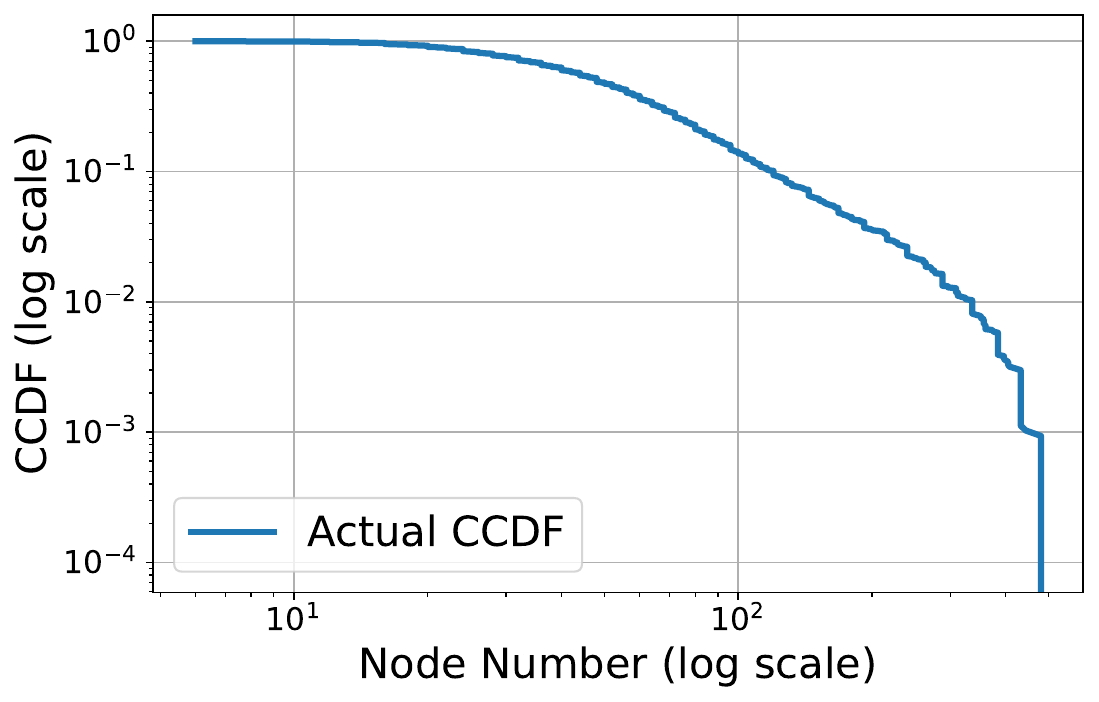}
        \label{}
    }
    \caption{Complementary Cumulative Distribution Function (CCDF) visualization on MetaModulus dataset.}
    \label{fig:ccdf}
\end{figure*}
\begin{figure*}[tb]
    \centering
    \vspace{-2em}
    \subfloat[Young's Modulus.]{
        \includegraphics[width=0.33\linewidth]{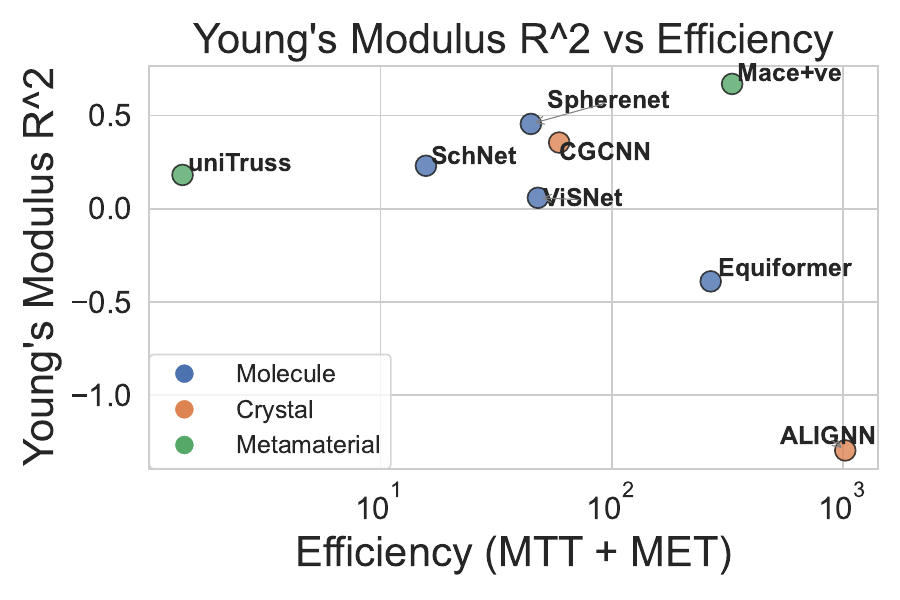}
        \label{fig:pred_young}
        }
    \subfloat[Shear Modulus.]{
        \includegraphics[width=0.33\linewidth]{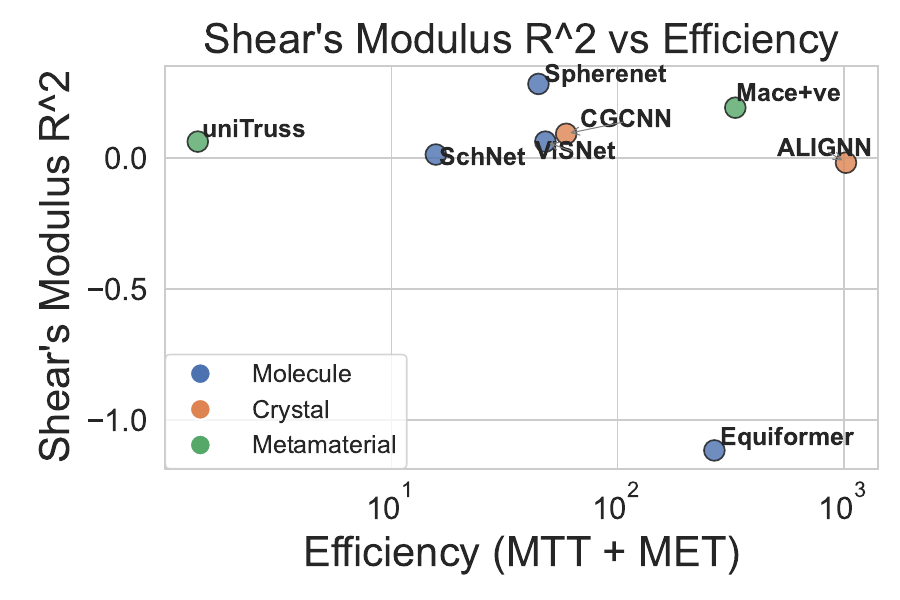}
        \label{fig:pred_shear}
    }
    \subfloat[Poisson's Ratio.]{
        \includegraphics[width=0.33\linewidth]{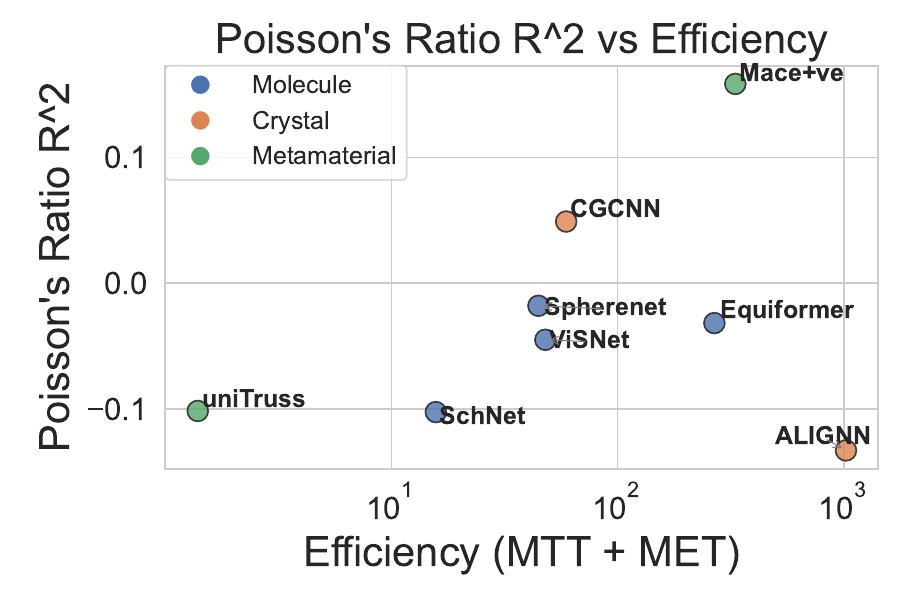}
        \label{fig:pred_poisson}
    }
    \caption{Comparisons of predictive models. Top-Left is better.}
    \vspace{-0.5em}
    \label{fig:bench_pred}
\end{figure*}
Moreover, to compare the node distributions, we compute Gini coefficient~\cite{wang2024towards} of the two datasets. The Gini coefficient reflects the long-tailedness distribution of a dataset, where a higher value indicates a more heterogeneous distribution~\cite{wang2024towards}. The results in Table~\ref{tab:app_sanit_modulus} reveal that the data sanitization process not only increases the validity ratios of the dataset but also reduces the heterogeneous degree of node distribution. In addition to the Gini coefficient, we provide a visualization of the Complementary Cumulative Distribution Function (CCDF) as shown in Figure~\ref{fig:ccdf}, from which it can be seen that the processed data contains fewer samples with large node numbers and the overall node number distribution becomes more homogeneous.

\subsection{Detailed Algorithm Comparisons}
\begin{figure}[tb]
    \centering
    \vspace{-1em}
    \includegraphics[width=0.8\linewidth]{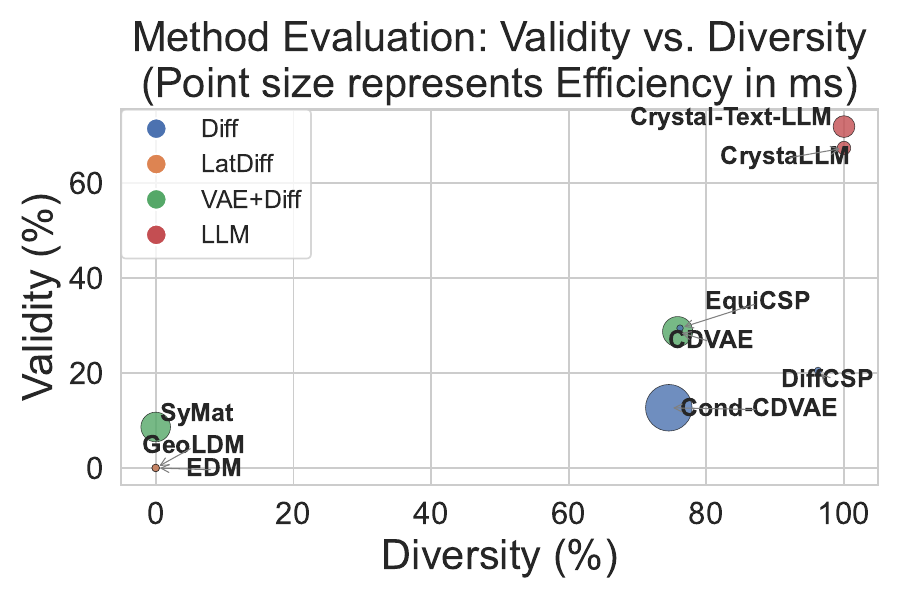}
    \caption{Comparisons of generative models. Top-Right-Small is better.}
    \label{fig:bench_gen}
    \vspace{-1em}
\end{figure}

Figures~\ref{fig:bench_pred} and \ref{fig:bench_gen} visualize the comparisons of benchmarks on generative task and predictive task, respectively.
\textbf{For predictive models (Figure~\ref{fig:bench_pred})}, we can summarize the following conclusions: (1) Mace+ve is the strongest predictive model across all the three mechanical properties, but its efficiency cannot be balanced. (2) Generally, the approaches with invariant constraints (SchNet, SphereNet, and CGCNN) are more efficient than the approaches with equivariant constraints (Mace+ve, ViSNet, and Equiformer). (3) ViSNet, Spherenet, and CGCNN are competitive for predicting three mechanical properties and perform balance regarding both efficiency and accuracy.
\textbf{For generative models (Figure~\ref{fig:bench_gen})}, we have the following observations: (1) LLM-based models, \ie, Crystal-Text-LLM and CrystaLLM are in top-right region, which indicates the superior performance on generative task \wrt\ proposed validity and diversity metrics. (2) DiffCSP performs well regarding diversity and exhibits good efficiency compared to other models. However, the validity of its generated metamaterial lattices cannot be guaranteed. (3) EquiCSP and CDVAE are balanced in validity, diversity, and efficiency. (4) The two molecular-targeted models, \ie, GeoLDM and EDM, are not suitable for metamaterials. We speculate one reason is the lack of periodicity constraints.

\end{document}